%% file: main.tex
\documentclass[submission, Phys]{SciPost}
\input{incl_settings.tex} 
\input{incl_shortcuts.tex}

\graphicspath{{./figs/}}

\begin{document}

\begin{flushright}
    {DESY-25-098}
\end{flushright}

\begin{center}{\LARGE \textbf{
\cp-Analyses with Symbolic Regression 
}}\end{center}

\begin{center}
  Henning Bahl\textsuperscript{1}, 
  Elina Fuchs\textsuperscript{2,3,4}, 
  Marco Menen\textsuperscript{2,3}, and
  Tilman Plehn\textsuperscript{1,5}
\end{center}

\begin{center}
{\bf 1} Institut f\"ur Theoretische Physik, 
Universit\"at Heidelberg, Philosophenweg 16, 69120 Heidelberg, Germany\\
{\bf 2} Institut f\"ur Theoretische Physik, 
Leibniz Universit\"at Hannover, Appelstraße 2, 30167 Hannover, Germany\\
{\bf 3} Physikalisch-Technische Bundesanstalt, Bundesallee 100, 38116 Braunschweig, Germany\\
{\bf 4} Deutsches Elektronen-Synchrotron DESY, Notkestr. 85, 22607 Hamburg, Germany \\
{\bf 5} Interdisciplinary Center for Scientific Computing (IWR), Universit\"at Heidelberg, Germany 
\end{center}

%\begin{center}
%\today
%\end{center}

% For convenience during refereeing: line numbers
%\linenumbers

\section*{Abstract}
         {\bf 
         Searching for \cp violation in Higgs interactions at the LHC is as challenging as it is important. Although modern machine learning outperforms traditional methods, its results are difficult to control and interpret, which is especially important if an unambiguous probe of a fundamental symmetry is required.
         We propose solving this problem by learning analytic formulas with symbolic regression. Using the complementary PySR and SymbolNet approaches, we learn \cp-sensitive observables at the detector level for WBF Higgs production and top-associated Higgs production. We find that they offer advantages in interpretability and performance.
         }

%\vspace{1pt}
\newpage
\noindent\rule{\textwidth}{1pt}
\tableofcontents\thispagestyle{fancy}
\noindent\rule{\textwidth}{1pt}
\vspace{1pt}

\clearpage
%%%%%%%%%%%%%%%%%%%%%%%%%%%%%%%%%%%%%%%%%%%%%%%%%%%%%%%%%%%%%%%%%%%
\section{Introduction}
\label{sec:intro}

More than ten years after the discovery of a SM-like Higgs boson~\cite{ATLAS:2012yve,CMS:2012qbp}, many questions about its nature remain unanswered~\cite{Dawson:2018dcd}. A particularly interesting property of the Higgs boson is its \cp nature. Any deviation from a fully \cp-even coupling of the Higgs to other SM particles would constitute a clear signal of BSM physics and could give us insight into how the matter-antimatter asymmetry observed in the Universe is generated~\cite{Sakharov:1967dj,Cohen:1993nk}. Although the rapidly growing LHC datasets allow for increasingly precise measurements, they also pose challenges for experimental analyses. First, data analysis techniques need to process huge datasets in a reasonable amount of time. At the same time, they have to work with sparse amounts of data to identify rare processes against the large background and study their properties. Finally, analyses need to cover large portions of phase space and BSM parameter space.

In recent years, various modern machine learning (ML) approaches have been developed to address these problems, offering an excellent compromise between speed, accuracy, and data efficiency.
One goal is to construct an observable that gives optimal sensitivity~\cite{Atwood:1991ka,Davier:1992nw,Diehl:1993br} to, for instance, \cp violation in the Higgs sector. At parton level, the Neyman-Pearson lemma~\cite{Neyman:1933wgr} gives an exact definition of the optimal observable. However, it is much harder to define once we include parton showering, detector resolution, and uncertainties in the reconstruction algorithms~\cite{Brehmer:2019bvj}. This issue can be approached using ML methods with very convincing results~\cite{Brehmer:2017lrt,Bhardwaj:2021ujv,Butter:2022vkj,Castro:2022zpq,Heimel:2023mvw,Bahl:2023qwk,Bortolato:2020zcg}. These powerful numerical methods do not lead to an analytic expression, so formulas for reco-level optimal observable are only known in rare cases, such as $p_{T,j_1} p_{T,j_2} \sin ( \Delta \phi_{jj} )$ in the context of VBF production~\cite{Butter:2021rvz}.

In addition to pure performance, the interpretability and control of ML-approaches is important in particular when it comes to probing fundamental symmetries. Interpretability can be achieved in a number of ways. For example, Shapley values allow to extract the relative importance of input parameters on ML-outputs~\cite{shapley,SHAP,Bahl:2023qwk}. However, they fall short of analytic equations. Symbolic regression (SR) allows us to extract the shape of a function and its parameters from data. In addition to theoretical insights, a single equation is fast to evaluate, and it simplifies reinterpretation. It is especially interesting in the context of symmetries as their conservation --- or violation --- will be imprinted in the function. In spite of its potential, SR has only been sparsely used in LHC physics so far~\cite{Butter:2021rvz,Dong:2022trn,Alnuqaydan:2022ncd,Tsoi:2023isc,AbdusSalam:2024obf,Morales-Alvarado:2024jrk,Soybelman:2024mbv,Tsoi:2024pbn}.

In this work, we show how SR can be used for detecting \cp violation in the Higgs sector. In particular, we demonstrate how it can be used to construct optimal \cp-odd observables and to reconstruct parton-level \cp-sensitive observables from reconstruction-level data. Throughout the paper, we compare results derived using two complementary SR-approaches. First, we use \pysr~\cite{cranmer2023interpretablemachinelearningscience}, which is based on the concept of evolutionary algorithms. Second, we use an enhanced version of \snet~\cite{Tsoi:2024ypg}, which relies on a sparsely connected neural network. After training, the NN resembles a fixed analytic equation that can be extracted.
Alternative tools include \texttt{AI Feynman} with its improved reconstruction thanks to the recursive simplification and 
\texttt{LASR}~\cite{Grayeli:2024LearnedConceptLibrary}, which uses large language models to accelerate SR.

In \cref{sec:symbolic_regression}, we first introduce the concept of SR and the functionalities of \pysr and \snet in detail. Then, we employ both algorithms to find an optimal \cp-odd observable in VBF Higgs production in \cref{sec:vbf_cp}. Afterwards, \cref{sec:ttH} deals with reconstructing the \cp-sensitive Collins-Soper angle $\cos \theta^*$ with an analytic expression at the reconstruction level. We conclude our findings in \cref{sec:conclusions}. 

%%%%%%%%%%%%%%%%%%%%%%%%%%%%%%%%%%%%%%%%%%%%%%%%%%%%%%%%%%%%%%%%%%%
\section{Symbolic regression}
\label{sec:symbolic_regression}

At the core of physics stands the claim that observations can be understood mathematically. For a set of data points from an experimental measurement, there has to be a function $f(x)$ that describes the distribution of those points. In some cases, the approximate form of $f(x)$ is known. For example, decaying particles are described by 
\begin{align}
    N(t) = a \cdot e^{-bt} +c \; .
\end{align}
The unknown parameters $a, b, c$ can be inferred from the data. 

In the absence of a priori information about $f(x)$ this method fails. On the numerical side, neural networks and on the analytic side symbolic regression (SR) do not assume a specific class of functions. Both determine a general function from a minimization problem
\begin{align}
    \hat{f} = \underset{f \in \mathcal{F}}{\arg \min}~\loss\Big(Y, f(\mathbf{X}) \Big) \; ,
\label{eq:symbolic_regression}
\end{align}
where $Y \in \mathbb{R}^{N_\text{data}}$ ($Y \in [0,1]^{N_\text{data}}$) are the labels in a regression (classification) task with $N_\text{data}$ data points, and $\mathbf{X} \in \mathbb{R}^{N_\text{data} \times N_f}$ are the inputs with $N_f$ features. For SR, $\mathcal{F}$ is the function space spanned by a set of basic operators defined by the user. 

Solving \cref{eq:symbolic_regression} is a challenging, NP-hard problem~\cite{virgolin2022symbolicregressionnphard}. Advanced methods are needed to determine $\hat{f}$ in a reasonable amount of time. SR algorithms have been shown to be able to tackle this problem for example by inferring well-known astrophysical formulas~\cite{Matchev:2021exoplanet,lemos2022rediscoveringorbitalmechanicsmachine,Khoo_2023} and in cosmology~\cite{Cranmer:2020wew}. In the following, we compare two different SR-approaches, \pysr and  \snet, for different tasks.

%%%%%%%%%%%%%%%%%%%%%%%%%%%%%%%%%%%%%%%%%%%%%%%%%%%%%%%
\subsection{\pysr}
\label{subsec:pysr}

\pysr~\cite{cranmer2023interpretablemachinelearningscience} is a Python module for symbolic regression that builds on an evolutionary algorithm and performs multiple evolutions at once. The formula finding happens via an inner and an outer loop. The inner loop corresponds to a single evolutionary algorithm with some modifications to speed up the training and improve the performance. The outer loop consists of a training part, where multiple populations evolve independently, and a migration part, in which the populations exchange expressions. 

%--------------------------------
\begin{figure}[b!]
    \centering
    \includegraphics[width=0.6\textwidth]{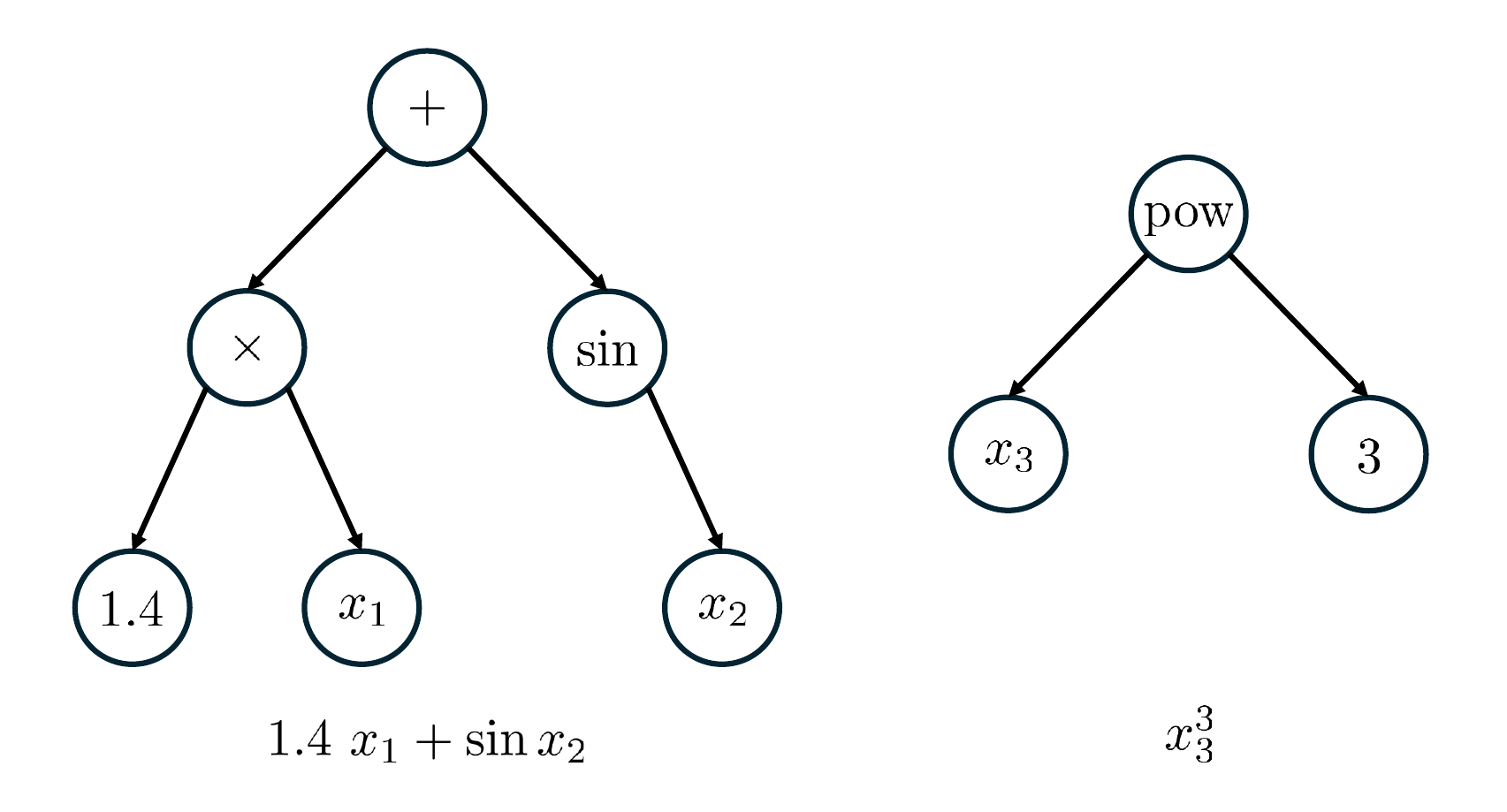}
    \caption[\pysr function tree]{Exemplary function trees displaying the functions $1.4~x_1 + \sin{x_2}$ and $x_3^3$ in \pysr.}
    \label{fig:pysr_tree}
\end{figure}
%--------------------------------

%--------------------------------
\begin{figure}[t]
    \centering
    \includegraphics[width=0.6\textwidth]{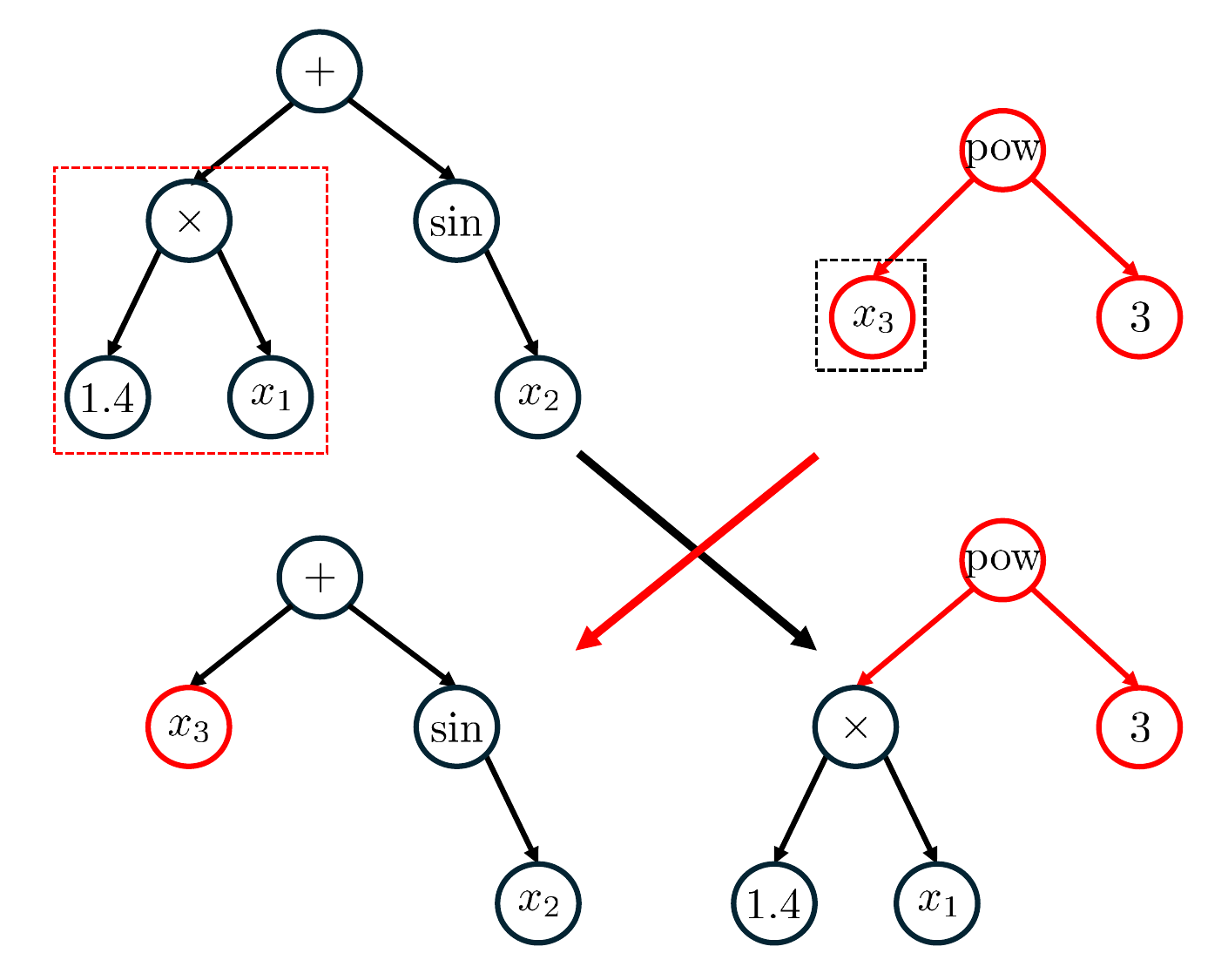}
    \caption[\pysr crossover operation]{A crossover operation in which the red-dashed part of the original tree has been exchanged with the black-dashed part of another tree. As a result the formulas are modified to $x_3 + \sin{x_2}$ and  $(1.4~x_1)^{3}$.}
    \label{fig:pysr_crossover}
\end{figure}
%--------------------------------

For the evolutionary algorithm defining the inner loop, formulas are represented by trees. Two simple examples can be seen in \cref{fig:pysr_tree}. The outermost leaves are variables and constants, combined by mathematical operations. These consist of unary and binary operations, either modifying a single leaf or combining two leaves, respectively. 

A population consists of a number of formula trees. Modifications to a single member of the population happen via tournament selection, where a subset of trees is compared by fitness. The fitness is determined by an objective function acting on the data. A single tree is chosen with a probability proportional to its fitness, to be cloned and modified in the next step. Modifications can include optimization of constants, simplification, mutation of a single node, and a crossover operation. In the latter, two trees exchange part of their structure, as shown in \cref{fig:pysr_crossover}. The new formula tree replaces the one with the lowest fitness in the population.

In contrast to other evolutionary algorithms, \pysr can use simulated annealing, where modifications of the trees can be rejected. A mutation is accepted according to the probability
\begin{align}
    p = \exp \frac{L_f-L_i}{\alpha T}  \; .
\label{eq:pysr_annealing}
\end{align}
Here, $L_f - L_i$ is the fitness difference from the modification, $T \in [0,1]$ is the temperature, varied during training, and $\alpha$ is a hyperparameter controlling the strength of the annealing. We use a modified version of \cref{eq:pysr_annealing}~\cite{Butter:2021rvz}, where the rejection probability reads
\begin{align}
    p = \exp \frac{L_f-L_i}{\alpha T L_i} \; .
\label{eq:pysr_norm_evol}
\end{align}
It does not reject functions with a better shape, but poorly initialized. 

Furthermore, \pysr alternates between evolution phases during which trees are primarily mutated and those during which they are primarily optimized. The mutation phase diversifies the population and finds expressions requiring an intermediate state, which would otherwise be simplified. The optimization phase improves the performance of the algorithm. Recently, \pysr introduced an adaptive parsimony. The parsimony parameter punishes the complexity of the formulas via a regularization term. Adaptive parsimony instead punishes complexities appearing more often in the population. This forces the algorithm to learn formulas of all complexities and explore a broader range of functions. Because the evolution of a population is achieved via the inner loop. the training of multiple populations can be parallelized.

In the outer loop, after a fixed number of iterations, the populations communicate and migrate members between them. The overall fittest members of each complexity are stored in the hall of fame. From here, expressions can also migrate to the populations. At the end of the training, the formulas stored in the hall of fame are returned. The final formula is determined by selection criteria such as the highest score evaluated by \pysr, or the lowest overall loss.

\pysr is powerful and highly customizable, outperforming other algorithms designed for similar problems~\cite{cranmer2023interpretablemachinelearningscience}. Its limitations appear for a large number of input features, or when a formula of high complexity is needed. Here, evolutionary algorithms struggle to converge within a limited time. Furthermore, the tree representation cannot be straightforwardly extended to other structures, like 4-vectors. 

%%%%%%%%%%%%%%%%%%%%%%%%%%%%%%%%%%%%%%%%%%%%%%%%%%%%%%%%%%%%%%%%%%%
\subsection{\snet}
\label{subsec:symbolnet}

The fact that evolutionary algorithms do not perform well for high-dimensional parameter spaces motivates the use of backpropagation. \snet~\cite{Tsoi:2024ypg} starts with a fully connected multi-layer perceptron (MLP) and replaces the activation functions with mathematical operators. They either modify a single node or combine nodes. The hidden layers are dubbed symbolic layers. During the so-called sparsity training the mathematical operators and the connections between nodes can be pruned to create a sparsely connected network. This prevents overtraining and simplifies the output formula for improved interpretability.

We change \snet in two ways. First, we adapt it to support 4-vectors by adding a vector dimension to the nodes. All operations are applied to the 4-vectors element-wise, and the vector dimension is collapsed in a specialized symbolic layer. This vectorized \snet requires extra checks because some mathematical operations can turn the 4-vectors unphysical. Second, we divide the training into three steps, to prevent gradient instabilities: 
\begin{itemize}
    \item default training, where only the usual NN weights and biases are trainable;
    \item mixed training, where every non-auxiliary parameter is trainable~\cite{Tsoi:2024ypg}; and
    \item sparsity training, only allowing for pruning of operators and connections.
\end{itemize}

The core of \snet are two \texttt{TensorFlow} custom layer classes, the input layer and the symbolic layer. They come with trainable and untrainable weights and thresholds, which are summarized in \cref{tab:snet_weights} and are explained in the following. 

The input features $\mathbf{X}_{\text{in}}$ are associated either with an auxiliary weight vector or with a threshold vector. The weight vector entries are not trainable and fixed to one. The threshold vector is clipped to the range $[0,1]$ and trainable during sparsity training. The input layer prunes input features which are not relevant for the training via
\begin{align}
        \mathbf{X}_{\text{in}} \to \mathbf{X}_{\text{in}} \theta(\mathbf{1} - \mathbf{T}_{\text{in}}) \; .
\end{align}
Without sparsity training, the input layer simply passes the input to the first symbolic layer.

%--------------------------------
\begin{figure}[t]
    \centering
    \includegraphics[width=0.6\linewidth]{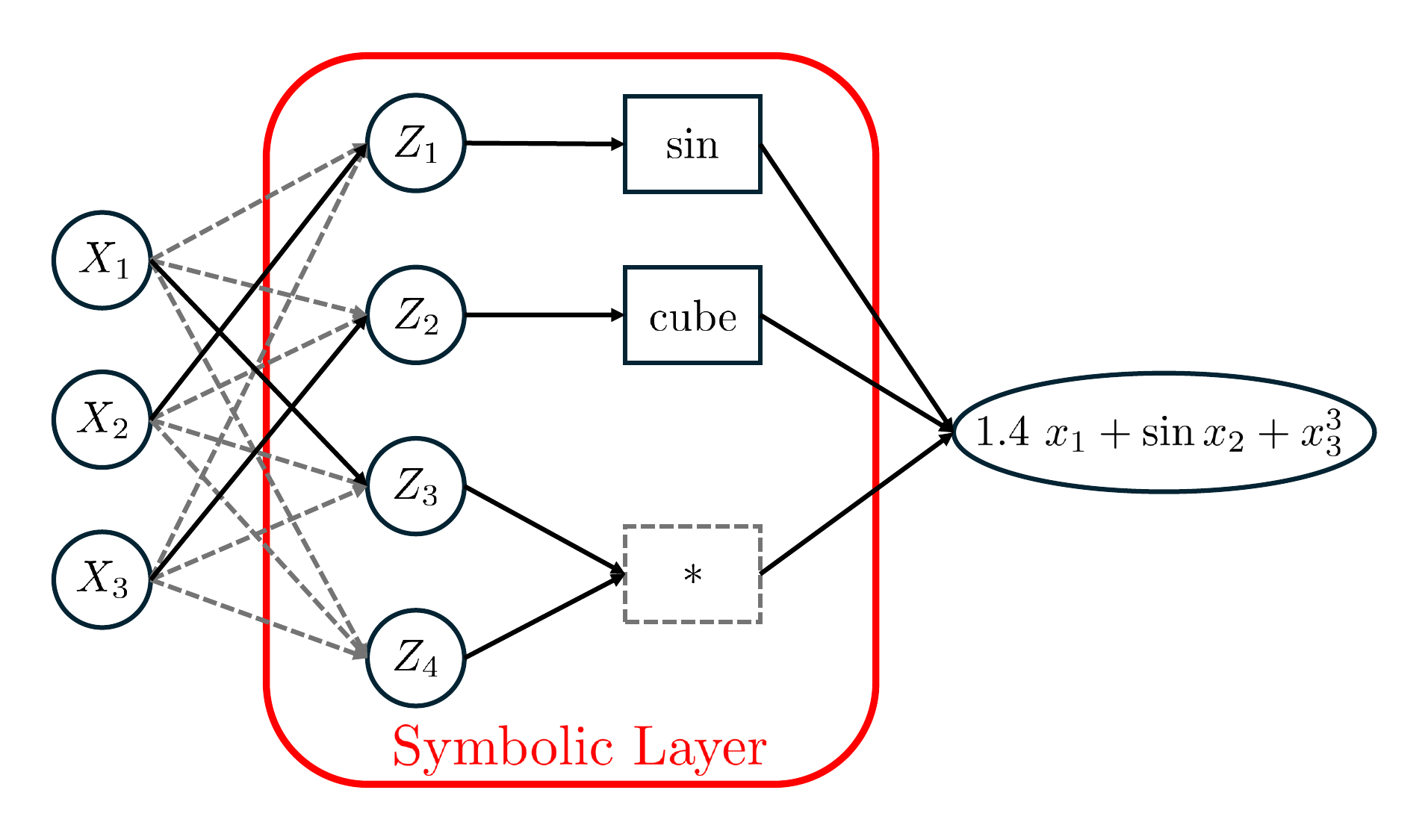}
    \caption{\snet architecture with a single symbolic layer. The input parameters are linearly transformed to an input representation for the symbolic layer. The mathmatical operations are applied in order, and the output of the symbolic layer is linearly transformed to the final prediction.
    }
    \label{fig:symbolic_layer}
\end{figure}
%--------------------------------

%--------------------------------
\begin{table}[b!]
    \centering
    \setlength{\arraycolsep}{5pt}
    \renewcommand{\arraystretch}{1.4}
    \begin{small} 
    \begin{tabular}{c|ccc|cc|c}
        \toprule
        Layer & Description & Label & Dim. & Values & Trainable & Dim. of trafo \\
        \midrule
        \multirow{2}{*}{Input layer} & Input weights & $\mathbf{W}_{\text{in}}$ & $n_{\text{in}}$ & $1$ & $\times$ & \multirow{2}{*}{\shortstack{$n_{\text{in}} \times l$ \\ $\to n_{\text{in}} \times l$}} \\
        & Input thresholds & $\mathbf{T}_{\text{in}}$ & $n_{\text{in}}$ & $[0,1]$ & M, S & \\
        \midrule
        \multirow{4}{*}{\shortstack{Linear trafo \\ in symbolic layer}} & Weights & $\mathbf{W}$ & $n \times m \times l$ & $\mathbb{R}$ & D, M & \multirow{4}{*}{\shortstack{$ n \times l$ \\ $\to m \times l$}} \\
        & Weight thresholds & $\mathbf{T}_{W}$ & $n \times m \times l$ & $\mathbb{R}^+$ & M, S &  \\
        & Biases & $\mathbf{B}$ & $m \times l$ & $\mathbb{R}$ & D, M & \\
        & Bias thresholds & $\mathbf{T}_{B}$ & $m \times l$ & $\mathbb{R}^+$ & M, S & \\
        \midrule
        \multirow{4}{*}{\shortstack{Symbolic trafo \\ in symbolic layer}} & Unary weights & $\mathbf{W}_{\text{unary}}$ & $n_f$ & 1 & $\times$ & \multirow{4}{*}{\shortstack{$m \times l$ \\ $\to k \times l'$}} \\
        & Unary thresholds & $\mathbf{T}_{\text{unary}}$ & $n_f$ & $[0,1]$ & M, S & \\
        & Binary weights & $\mathbf{W}_{\text{binary}}$ & $n_g$ & 1 & $\times$ & \\
        & Binary thresholds & $\mathbf{T}_{\text{binary}}$ & $n_g$ & $[0,1]$ & M, S & \\
        \bottomrule
    \end{tabular} 
    \end{small}
    \caption{Parameters used for the training of \snet alongside their dimension, possible values of their components, and the phases in which they are trainable. $n_f$ and $n_g$ are the number of unary and binary operators in the layer, while $m = n_f + 2n_g$ and $k = n_f + n_g$. $l,l' = \{1,4\}$, depends on the type of symbolic layer. D, M, and S stand for the training phases: default, mixed, and sparsity.}
    \label{tab:snet_weights}
\end{table}
%--------------------------------

Symbolic layers are split into two parts. The first one is a linear combination of the incoming features associated with the usual weights and biases of an MLP, complemented by threshold matrices. The default weights are initialized using a random uniform distribution, while the thresholds are initialized to zero. During default training, only the weights and biases are trainable, while in the sparsity setup, only the thresholds are. In the mixed setup, all four parameters are trainable. Pruning is implemented as
\begin{align}
        w_{ij} &\to w_{ij} \theta(|w_{ij}| - t_{w,ij})\;, \notag \\
        b_{i}  &\to b_{i} \theta(|b_{i}| - t_{b,i})\;.
\end{align}
In the second step, the output of the linear operation is passed to a set of pre-defined mathematical operations. This replaces the usual MLP activation. The operations are split into unary and binary operators. Similarly to the input features, each unary (binary) operator is associated with an untrainable auxiliary weight and a threshold, where the latter is trainable during sparsity training. They are pruned via 
\begin{align}
        f_i(a) &\to f_i(a) \; \theta(1 - t_{\text{unary},i}) + a \left(1 - \theta(1 - t_{\text{unary},i})\right) \notag \\
        g_i(a,b)& \to g_i(a,b) \; \theta(1 - t_\text{binary},i) + (a+b)\left(1 - \theta(1 - t_\text{binary},i)\right) \; .
\end{align}
Unary operators are simplified to the identity if the threshold parameter exceeds one, while binary operators are simplified to an addition. In the special case of a symbolic layer that takes vectors as its input and returns scalars, the pruning implies masked operators return zero,
\begin{align}
        f_i(a) &\to f_i(a) \; \theta(1 - t_{\text{unary},i}) \notag \\
        g_i(a,b)& \to g_i(a,b) \; \theta(1 - t_\text{binary},i) \; .
\end{align}

\begin{figure}
    \centering
    \includegraphics[width=0.7\linewidth]{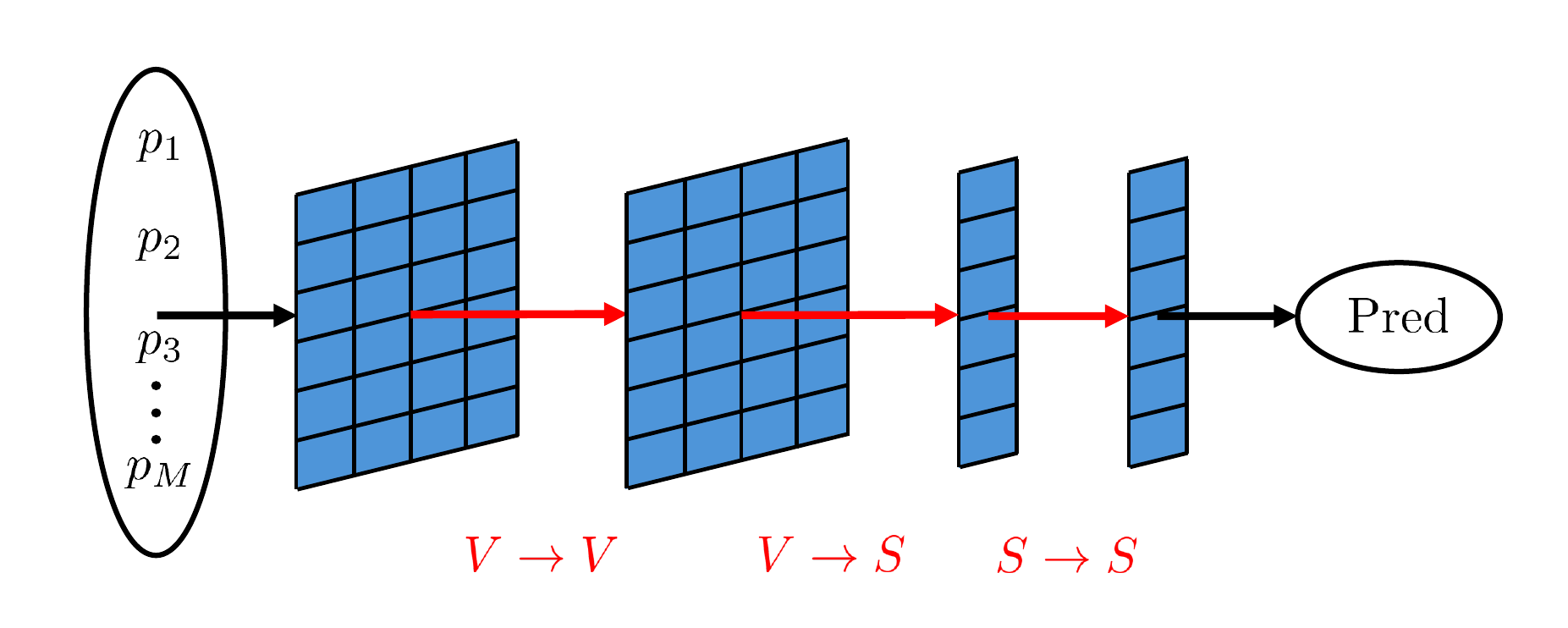}
    \caption{illustration of a vectorized \snet structure. The input $4$-vectors are transformed via one of each layer types to the final prediction.}
    \label{fig:symbolnet_structure}
\end{figure}

For our vectorized \snet, we split the symbolic layers into three types, depending on their input and output dimension, and illustrated in Fig.~\ref{fig:symbolnet_structure}.
\begin{itemize}
    \item $V\to V$ layers:

    Input and output of the symbolic layer are 4-vector-like objects. All operations are applied element-wise, with the exception of the Lorentz boost 
    \begin{align}
        g(p_i, p_j) \in \{ \mathrm{boost}(p_i | p_j), \ldots \} \;,
    \end{align}
    which boost the vector $p_i$ to the frame $p_j$ and which is unique to the $V\to V$ layer. There is no restriction to the number of $V\to V$ layers in \snet.

    \item $V\to S$ layer:

    Next, the 4-vector-like objects are transformed to scalars. Possible unary operators are 
    \begin{align}
        f(p) \in \left\{ p_0, p_x, p_y, p_z, \Vert p \Vert_3, \Vert p \Vert_4 \right\} \;. 
    \end{align}
    The $p_i$ pick a component of the vector, $\Vert p \Vert_3$ computes the Euclidian three-vector norm, and $\Vert p \Vert_4$ the Minkowskian norm. The binary operators consist of the three-vector and Minkowski products
    \begin{align}
        g(p_i,p_j) \in \left\{ \langle p_i \times p_j \rangle_3, \langle p_i \times p_j \rangle_4 \right\} \; .
    \end{align}
    Exactly one $V\to S$ layer is required in the vectorized \snet. All operations in this layer are unique to it. 

    \item $S\to S$ layers:

    They correspond to the default \snet layers and only work with scalar quantities. Any number of these layers may be used.

    \item Output layer:

    The final symbolic layer in the network is an $S\to S$ layer that only consists of a linear combination $\mathbb{R}^n\to\mathbb{R}$ and no further unary or binary operations.
\end{itemize}

%--------------------------------
%--------------------------------

During backpropagation, the step function responsible for masking the operations is replaced by the derivative of the sigmoid function 
\begin{align}
    \frac{d\theta(x)}{dx} \simeq \frac{\kappa e^{-\kappa x}}{(1+e^{-\kappa x})^2}
    \qquad \text{with} \quad \kappa = 5 \; .
\end{align}
The training loss is 
\begin{align}
    \loss = \loss_\text{base} + \loss_\text{sparse} \; ,
\end{align}
where $\loss_\text{base}$ is the default loss, \ie Mean Squared Error (MSE) or the cross-entropy. The additional sparsity loss controls the dynamic pruning and is set to zero during the default training. It is given by
\begin{align}
  \loss_\text{sparse}
  &= \loss_\text{error} D(s_\text{weight};\alpha_\text{weight},d) \loss_\text{threshold}^\text{weight} \notag \\
  &+ \loss_\text{error} D(s_\text{aux};\alpha_\text{aux},d) \loss_\text{threshold}^{\text{aux}}\;.
\end{align}
The target sparsities and the aggressiveness of the pruning are controlled via the decay factor
\begin{align}
    D(s;\alpha,d) = \exp\left[-\left(\frac{\alpha}{\alpha - \min(s,\alpha)}\right)^d + 1\right] \; ,
\end{align}
where $s$ are the current sparsity values, $\alpha$ the target sparsities set by the user, and $d$ is controlling the aggressiveness of the pruning. We use the default choice $d=0.01$. 
To encourage large threshold values, we use
\begin{align}
    \loss_\text{threshold}^\text{weight} &= \frac{1}{n_\text{weight}}\sum_{i} e^{- t_{w,i}} \\
    \loss_\text{threshold}^{\text{aux} =\{\text{input,unary,binary}\}} &= \exp\left(\frac{1}{n_\text{aux}}\sum_i^{n_\text{aux}}t_{\text{aux},i}\right) \; .
\end{align}

%%%%%%%%%%%%%%%%%%%%%%%%%%%%%%%%%%%%%%%%%%%%%%%%%%%%%%%
\section{\texorpdfstring{\cp}{CP}-odd observables for WBF-Higgs production}
\label{sec:vbf_cp}

In order to conclusively test \cp for example in WBF Higgs production, one needs to learn and employ the optimal \cp-odd observable for this process. The standard theory framework for this question is the dimension-6 SMEFT operator
\begin{align}
    \frac{\cHWtil}{\Lambda^2}\varphi^\dagger\varphi\widetilde W_{\mu\nu}^I W^{I\mu\nu} \; ,
\end{align}
where $\varphi$ is the Higgs doublet, $W$ the $SU(2)_L$ field strength, $\widetilde W$ its dual, and $I$ the $SU(2)_L$ index. Two more dimension-6 operators induce \cp violation in WBF Higgs production at the tree level,
\begin{align}
 \varphi^\dagger\varphi\widetilde B_{\mu\nu}B^{\mu\nu}
 \qquad \text{and} \qquad 
 \varphi^\dagger\tau^I\varphi\widetilde W_{\mu\nu}^IB^{\mu\nu} \; .
\end{align}
All three operators induce the same Lorentz structure in the $HZZ$ coupling, but only \cHWtil affects the $HWW$ coupling. This is why we can stick to \cHWtil for simplicity.
In this study, we consider the $H\to\gamma\gamma$ decay channel, taking into account also the effect of \cHWtil on the decay rate.

%%%%%%%%%%%%%%%%%%%%%%%%%%%%%%%%%%%%%%%%%%%%%%%%%%%%%%%%%%%%%%%%%%%
\subsection{Optimal \texorpdfstring{\cp}{CP}-observable}
\label{sec:training_cpodd_observable}

Considering a process with a contribution proportional to the \cp-violating coupling $\theta$, we can write the squared amplitude as
\begin{align}
    |\mathcal{M}|^2 = |\mathcal{M}_{\cpe}|^2 + 2\theta \text{Re}\left[\mathcal{M}_{\cpe}\mathcal{M}_{\cpo}^*\right] + \theta^2 |\mathcal{M}_{\cpo}|^2\;. 
\label{eq:mat_fact}
\end{align}
The first and last terms are effectively \cpe, while \cp violation only appears through the interference. Correspondingly, we split the single-event likelihood into a \cp-even and a \cp-odd part,
\begin{align}
    p(x|\theta) = \frac{1}{\sigma(\theta)}\frac{d^d\sigma(x|\theta)}{dx^d} = p_e(x|\theta) + p_o(x|\theta) \; .
\label{eq:cpv_like}
\end{align}
Here $\sigma(\theta)$ is the total cross-section and $d^d\sigma/dx^d$ the differential cross section with respect to the observable $x$. The optimal \cp-odd observable for a given $\theta$ is~\cite{Davier:1992nw}
\begin{align}
    \omega_{\cpo} = \frac{p_o}{p_e}\,.
\end{align}
Using ML, a classifier converges to this optimal observable when trained to distinguish two samples with finite values $\pm \theta$. Using Eq.\eqref{eq:cpv_like} we can generate the two samples drawing from $p_e(x|\theta) \pm p_o(x|\theta)$, so the Neyman-Pearson-optimal classifier is
\begin{align}
\label{eq:tth_optimal_observable}
    D(x) 
    &= \frac{p_e(x|\theta) + p_o(x|\theta)}{p_e(x|\theta) + p_o(x|\theta) \; + \;  p_e(x|\theta) - p_o(x|\theta)} \notag \\
%tp    &=\frac{p_e(x|\theta) + p_o(x|\theta)}{2p_e(x|\theta)}  \notag \\
    &= \frac{1+\omega_\cpo(x)}{2} \notag \\
\Leftrightarrow \qquad 
    \omega_\cpo(x) &= 2 D(x) - 1\;.
\end{align}
This strategy can be applied for any classifier, including boosted decision trees (BDTs)~\cite{Castro:2022zpq}. 

Using symbolic regression has the advantage that we can check analytically if the learned observable is indeed \cp-odd. To define the classification task we assign the label `1' to the sample with positive \cHWtil and `0' to the sample with negative \cHWtil. The classifier output is mapped to the interval $[0,1]$ as
\begin{align}
    D(x) = \text{sigmoid}\left(d(x)\right)\;,
\end{align}
where $d(x)$ is the analytic expressions we target with symbolic regression. Since $2\,\text{sigmoid}(x) - 1$ is odd under $x\rightarrow -x$, the \cp-properties of $d$ and $\omega_\cpo$ are the same. 

To improve the training we add an additional term to the cross-entropy loss $\loss_\text{CE}$, penalizing non-\cp-odd observables,
\begin{align}
    \loss 
%    &= \loss_\text{CE} + \loss_\cp \notag \\
     &= \loss_\text{CE} + \frac{\alpha}{n_\cp} \sum_{i=1}^{n_\cp} \left[ d(x_i) + d(x^\cp_i)\right]  \;, 
\label{eq:cploss}
\end{align}
where $x^\cp_i$ is the \cp-transformed input. The sum runs over $n_\cp$ phase space points and $\alpha$ balances the two loss contributions. If $d$ is \cp-odd, then $d(x^\cp_i) = - d(x_i)$ and $\loss = \loss_\text{CE}$.

At this point a difference between \pysr and \snet becomes relevant: \pysr starts with simple expressions and generates more complex expressions over time. The additional \cp-odd loss of Eq.\eqref{eq:cploss} prevents the generation of non-\cp-odd formulas. While this prevents \cp-even equations as intermediary mutation steps, we find this to not affect performance. In contrast, \snet starts from a complicated expression and prunes it over time. The \cp-odd loss term does not necessarily result in \cp-odd formulas.

%%%%%%%%%%%%%%%%%%%%%%%%%%%%%%%%%%%%%%%%%%%%%%%%%%%%%%%%%%%%%%%%%%%
\subsection{Events and training}

We generate leading-order events for the hard process
\begin{align}
  pp \to H_{\gamma \gamma} jj 
\end{align} 
using \madgraph~3.5.0~\cite{Alwall:2011uj} with the \textsc{SMEFTsim} \textsc{UFO} model~\cite{Brivio:2017btx,Brivio:2020onw}. For the parton shower, we use
\pythia~\cite{Sjostrand:2014zea}; for the detector simulation,
\delphes~\cite{deFavereau:2013fsa}; and for the jet clustering,
\fastjet~\cite{Cacciari:2011ma}.
We generate data for 11 different scenarios,
\begin{align}
\cHWtil \in \left\{ 0, \pm 0.1, \pm 0.25, \pm 0.5, \pm 0.75, \pm 1 \right\}  \; .
\end{align}
Unless mentioned otherwise, we use 250k events for training and testing as well as 100k events for validation. For the analysis we require exactly two photons and at two tagging jets and impose the pre-selection cuts
\begin{alignat}{9}
  m_{\gamma \gamma} &= 110~...~140~\gev\;, &\qquad 
  \frac{p_{T,\gamma_{1,2}}}{m_{\gamma\gamma}} &> 0.35,0.25\;, \notag \\
  p_{T,j} &>30~\gev\;, &\qquad  
  |\eta_j| &< 4.4\;, &\qquad 
  \Delta\eta_{jj} &> 2 \; .
\end{alignat}
The Higgs 4-momentum is reconstructed from the two photons. 

Following Sec.~\ref{sec:training_cpodd_observable} we train our WBF \cp-odd observables through a classifier to distinguish events with positive and negative \cHWtil. 
We apply a sigmoid function to the learned formulas to ensure $D \in [0,1]$.
In our training, we use
\begin{align}
 \left\{ \quad 
 x_{j_{1,2}} = \frac{p_{T,j_{1,2}}}{m_h},
 \eta_{j_1}, \phi_{j_1},  \quad
 \Delta \eta_{jj}, \Delta \phi_{jj}, 
  x_{jj} = \frac{m_{jj}}{m_h}, \quad
  x_h = \frac{p_{T,h}}{m_h}, \eta_h, \phi_h 
  \quad \right\} 
\end{align}
for jets ordered in $p_T$ and with $m_h = 125\, \gev$.  

For \pysr, we train for 500 iterations with a maximum complexity of 60 using sine, cosine, absolute value, exponential, logarithm, sine hyperbolicus, cosine hyperbolicus, addition, multiplication, and division as operators. After training, we optimize the numerical constants.
As our \snet network, we use two symbolic layers with the functions sine, cosine, absolute value, square, square root, multiplication, and division. We here use non-vectorized symbolic layers since the number of input features is comparably small. We train the network for 1000 epochs with $\alpha_\text{input}=0.6$, $\alpha_\text{weight} = 0.6$, $\alpha_\text{unary} = \alpha_\text{binary} =0.3$. No afterburning is necessary, since the loss minimization already optimizes the numerical constants.

%%%%%%%%%%%%%%%%%%%%%%%%%%%%%%%%%%%%%%%%%%%%%%%%%%%%%%%%%%%%%%%%%%%
\subsection{Results and formulas}

In addition to the $\omega_\cpo$ distributions, we also compute the bin-wise asymmetry 
\begin{align}\label{eq:asym}
\mathcal{A}_i = \frac{N_i^+-N_i^-}{N_i^++N_i^-}
\qquad \text{with} \qquad 
N_i^+ &= \#[\omega_{\cpo,i}, \omega_{\cpo,{i+1}}] \notag \\
N_i^- &= \#[-\omega_{\cpo,i + 1}, -\omega_{\cpo,i}] \; ,
\end{align}
For the statistical uncertainty of the $\mathcal{A}_i$ we use an integrated  luminosity of $300\,\ifb$. Details on the calculation of the standard deviations can be found in App.~\ref{app:asymmetry}.

%--------------------------------
\begin{figure}[b!]
    \includegraphics[width=.99\textwidth]{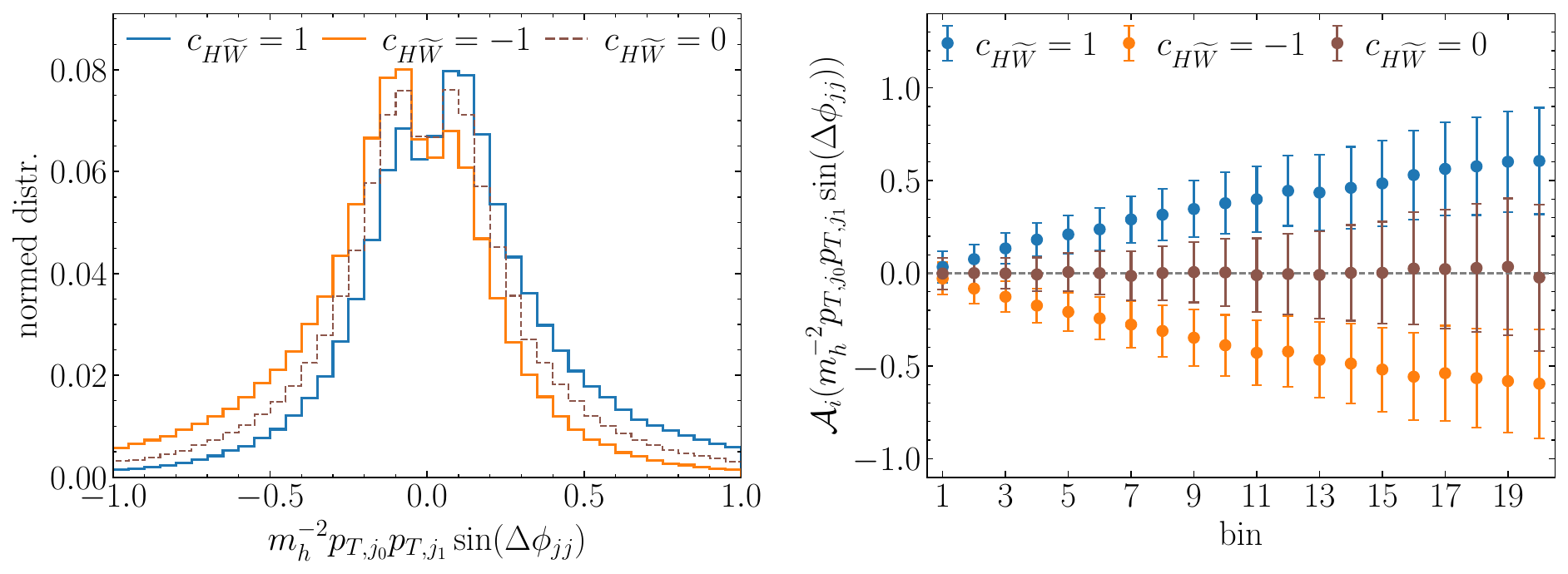}
    \caption{Normalized distribution of $p_{T,j_1}p_{T,j_2}\sin \Delta\phi_{jj}$ (left) and its bin-wise asymmetry $\mathcal{A}$ defined in Eq.~\ref{eq:asym} (right) for the SM (corresponding to $\cHWtil = 0$) and $\cHWtil = \pm 1$.}
    \label{fig:vbf_cpodd_deltaphi}
\end{figure}
%--------------------------------

In the left panel of Fig.~\ref{fig:vbf_cpodd_deltaphi} we show the observable 
\begin{align}
\frac{1}{m_h^2} p_{T,j_1}p_{T,j_2} \sin \Delta\phi_{jj} \; ,
\end{align}
the most sensitive \cp-odd observable at the parton level for small \cHWtil~\cite{Plehn:2001nj,Hankele:2006ma,Brehmer:2017lrt,Butter:2021rvz}. It provides a baseline for our symbolic regression. While the SM distribution is symmetric around zero, the distributions with \cp violation are asymmetric.

The relation between \cp and this bin-wise asymmetry is confirmed in the right panel of Fig.~\ref{fig:vbf_cpodd_deltaphi}. The SM distribution is symmetric, with small statistical fluctuations. For $\cHWtil = 1$ we see a positive asymmetry, while for $\cHWtil = -1$ the asymmetry also changes sign. The higher bins feature a larger asymmetry in comparison to the inner bins, but with a larger statistical uncertainty, suggesting that there will be a sweet spot for the analysis.

%--------------------------------
\begin{figure}[t]
    \includegraphics[width=.99\textwidth]{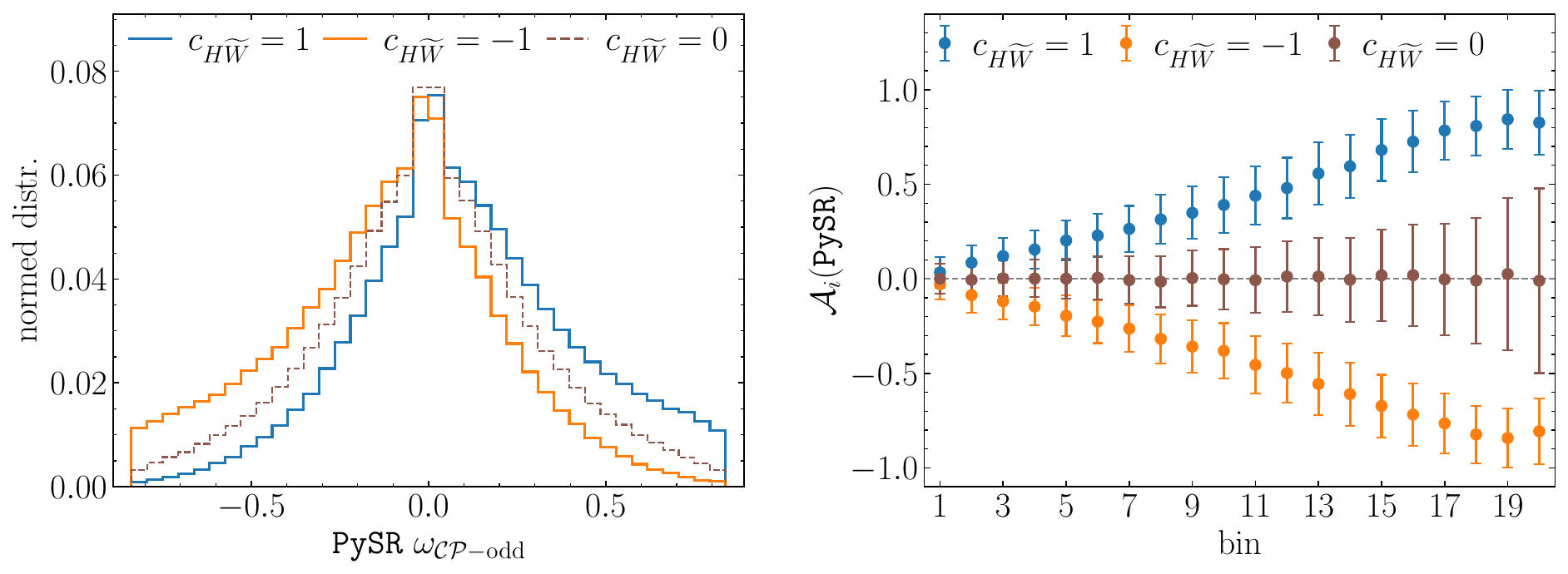} \\
    \includegraphics[width=.99\textwidth]{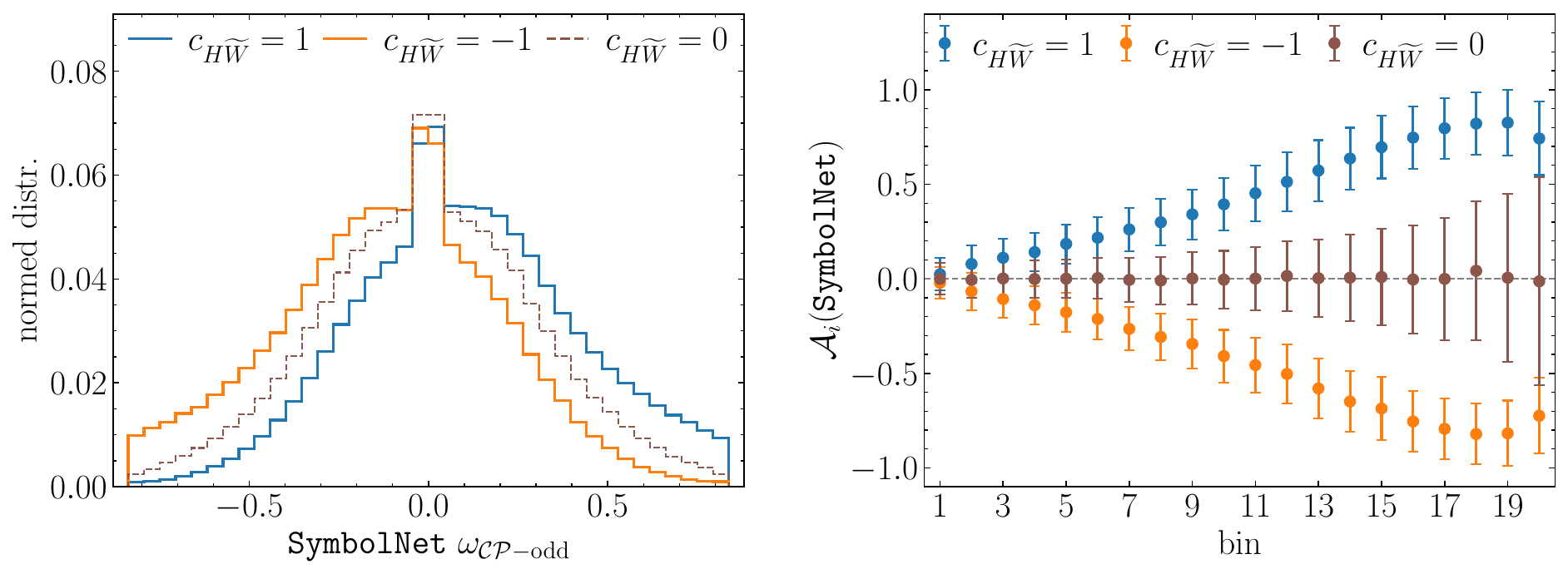}
    \caption{Performance of the learned formulas for the \cpo observable from \pysr and \snet (left) and their bin-wise asymmetries $\mathcal{A}$ (right).}
    \label{fig:vbf_cpodd_SR}
\end{figure}
%--------------------------------

Our learned formulas from \pysr and \snet are evaluated in the left panels of Fig.~\ref{fig:vbf_cpodd_SR}. Both are very similar, the \snet distribution being slightly wider than its \pysr counterpart.  Both reflect the \cpo nature, with a symmetric SM distribution and the asymmetric but mirrored outcomes for $\cHWtil = \pm 1$. Compared to $p_{T,j_1}p_{T,j_2}\sin \Delta\phi_{jj}$ in Fig.~\ref{fig:vbf_cpodd_deltaphi}, the \pysr and \snet distributions are wider and have less separation around zero.
The analysis power of the learned formulas is illustrated in the right panels of Fig.~\ref{fig:vbf_cpodd_SR}. Both formulas have the largest asymmetry for the highest bins, while the most significant bins are in the middle. 

Finally, we can confirm the \cpo nature of the learned formulas from the analytic forms,
\begin{align}
    d^{\pysr} &= \frac{1.8566 \, \textcolor{blue}{\sin \Delta\phi_{jj}}}{\left|\dfrac{0.3080x_{j_1}\log \Delta\eta_{jj}  + \log \Delta\eta_{jj} \sinh(x_{j_2} - 2.5977) + 0.3080\sinh x_h }{x_{j_1}\log \Delta\eta_{jj} + \sinh x_h }\right| + 0.6047} \notag  \\[2mm]
    d^{\snet} &= 0.715 \textcolor{blue}{\Delta\phi_{jj}} \Big[-0.348 (0.561 {x_{j_2}} - 0.25 {\Delta\eta_{jj}} + 0.0315 {x_{jj}} + 0.746 {x_{h}}) - 0.27\Big ]
    \notag \\
    &\hspace{2.0cm}\cdot \Big[ 0.0493 (0.603 {\Delta\eta_{jj}} - 0.0811 {x_{jj}} - {x_{h}})^2  \notag \\
    &\hspace{2.35cm} - 0.654 |0.463 {x_{j_2}} + 0.477 {\Delta\eta_{jj}} + 0.373 {x_{h}}|^{0.5} \notag \\
    &\hspace{2.35cm}- 0.134 \sin(-0.555 {\Delta\eta_{jj}} + 0.345 {x_{jj}} + 0.443 {x_{h}})  \notag \\
    &\hspace{2.35cm}- 1.82 \cos(0.642 {\Delta\phi_{jj}}) \Big]\;,
\end{align}
where we color the \cpo part, to confirm that the learned observables are actually \cpo. We observe that a simple \cpo structure is multiplied by a more complex \cpe function, retaining the \cpo nature of the overall expression.
While for \pysr, we explicitly enforce this by including the \cpo loss contribution, we do not include it in the \snet training. 
As we will see below, a \cpo loss destabilizes the training. 
By construction, the learned \snet formula does not have to be \cpo, and the formula learned in another \snet run --- using the same settings --- confirms this,
\begin{align}
    d^{\snet} ={}& -0.19 \Big[ 0.87 {\Delta\phi_{jj}} + 0.063 {x_{h}}\Big] \Big[ -0.66 {x_{j_2}} + 0.050 {\Delta\phi_{jj}} - 1.1 {x_{h}} - 0.02 {\phi_{h}} \Big] \notag \\
    & - 0.65 \sin\Big[0.49 {x_{j_2}} + 0.23 {\Delta\phi_{jj}} + 0.28 {x_{h}} \notag \\
    &\hspace{1.5cm} - (0.082 {\Delta\phi_{jj}} + 0.0059 {x_{h}}) (0.66 {x_{j_2}} - 0.05 {\Delta\phi_{jj}} + 1.1 {x_{h}} + 0.02 {\phi_{h}}) \notag \\
    &\hspace{1.5cm}  + (0.35 {\Delta\phi_{jj}} + 0.18 {x_{h}}) (-0.60 {x_{j_2}} - 0.66 {\Delta\phi_{jj}} - 0.36 {x_{h}}) \nonumber\\
    &\hspace{1.5cm} + 0.42 \cos(0.50 {x_{j_2}} - 1.3 {\Delta\phi_{jj}} + 0.46 {x_{h}})\Big] \notag \\
    & + 0.62 \cos \Big[ 0.50 {x_{j_2}} - 1.3 {\Delta\phi_{jj}} + 0.46 {x_{h}} \Big] \;.
\end{align}
Even though it discriminates the SM data better from the $\cHWtil = \pm 1$ cases, it is not \cpo, so a deviation from the SM based on this observable is not a unique sign of \cp violation.

%--------------------------------
\begin{table}[b!]
    \centering
    \begin{small} \begin{tabular}{c|cc}
        \toprule
          & $\sigma(\cHWtil=1\text{ vs.\ SM})$ & $\sigma(\cHWtil=0.25\text{ vs.\ SM})$ \\
        \midrule
        $p_{T,j_1}p_{T,j_2}\sin\Delta\phi_{jj}$ & 6.76 & 2.43 \\
        \cmidrule(l){1-1}
        trained on $\cHWtil=\pm 1$ & & \\
        \cmidrule(l){1-1} 
        \pysr                           & 6.98 & 2.47 \\
        \snet                      & 7.07 & 2.49 \\
        BDT                                     & 6.71 & 2.36 \\
        \cmidrule(l){1-1}
        trained on $\cHWtil=\pm 0.1$ & & \\
        \cmidrule(l){1-1} 
        \pysr                           & 7.07 & 2.43 \\
        \snet                      & 1.67 & 0.82 \\
        BDT                                     & 3.27 & 1.26 \\
        \bottomrule
    \end{tabular} \end{small}
    \caption{Significances for distinguishing the dimension-6 hypothesis from the SM, evaluated using the various learned \cp-odd observables.}
    \label{tab:vbf_cp_odd_significances}
\end{table}
%--------------------------------

From the bin-wise asymmetry we can calculate significances for $300\ifb$, as detailed in App.~\ref{app:asymmetry}. These significances are listed in Tab.~\ref{tab:vbf_cp_odd_significances}. The significances for negative \cHWtil are identical to their positive counterparts.
For comparison, we also show significances from a numerical BDT, trained using \XGBoost~\cite{2016arXiv160302754C}. As we know from our two \snet formulas, the BDT observables can have \cpe components enhancing the signal significance. However, this does not imply a higher significance of discovering \cp violation. While we can check this artifact for the formulas from \pysr and \snet, this is not possible for a BDT or neural-network classifier.

The significances quoted for $\cHWtil = 1$ vs SM lie around $7\,\sigma$ for the case where the training data is with $\cHWtil = \pm 1$. The learned \snet and \pysr formulas slightly outperform the classic $p_{T,j_1}p_{T,j_2}\sin \Delta\phi_{jj}$. 
%The BDT performance is slightly worse. 
Training on $\cHWtil = \pm 0.1$, \ie with a much smaller \cpo contribution, the performance of the \snet and BDT observables drops. In contrast, the \pysr formula is robust to the less sensitive training data. For a weaker signal, $\cHWtil = 0.25$, but still trained on $\cHWtil = \pm 0.1$, the typical significances shrink to around $2.5\,\sigma$, but the performance pattern of the different approaches remains, \ie \pysr performs better than the BDT and \snet.

%--------------------------------
\begin{figure}[t]
    \centering
    \includegraphics[width=.50\textwidth]{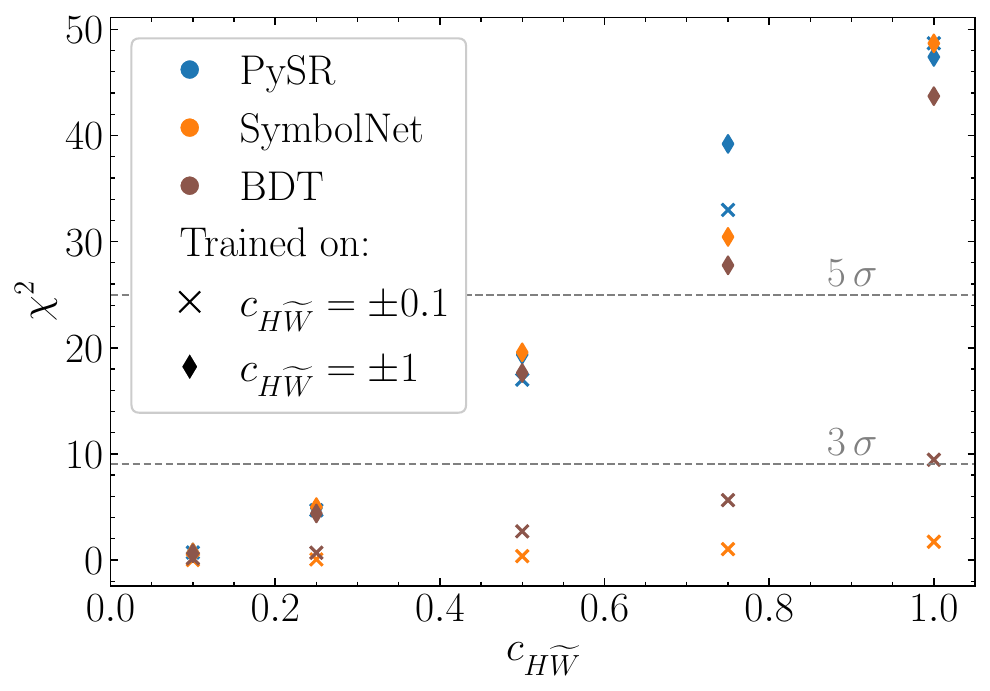}
    \caption{Dependence of the test significance on \cHWtil. For the diamonds, the observables are trained with $\cHWtil = \pm 1$, while for the crosses the observables are trained with $\cHWtil = \pm 0.1$. For the two lowest \cHWtil values, the blue crosses are hidden behind the diamond markers.
    }
    \label{fig:vbf_cpodd_chwtil_efficiency}
\end{figure}
%--------------------------------

%-------------------------------- 

In Fig.~\ref{fig:vbf_cpodd_chwtil_efficiency} we investigate the dependence of the performance of the learned formulas on the size of \cHWtil in the training data. It confirms the pattern observed in Tab.~\ref{tab:vbf_cp_odd_significances} ---  since \pysr builds formulas from simple to complicated, it is very efficient even for little \cp violation in the training data. The \pysr $\chi^2$ values are stable for formulas trained on $\cHWtil = \pm 0.1~...~\pm 1$. In contrast, the \snet and BDT performances suffer for less \cp violation in the training data.

%--------------------------------
\begin{figure}[tb]
    \centering
    \includegraphics[width=0.55\textwidth]{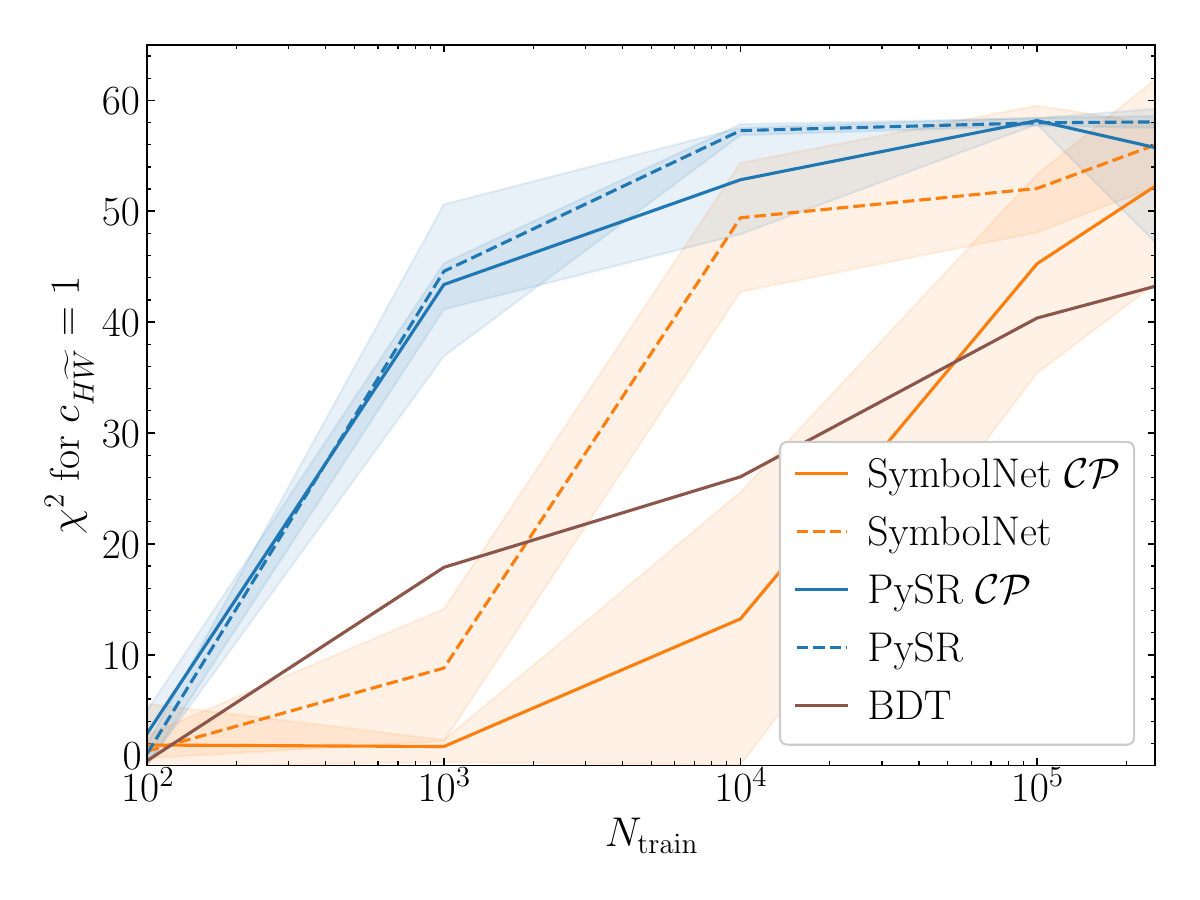}
    \caption{Test significance as a function of the training statistics. The label \cp indicates the additional \cp-odd loss of Eq.\eqref{eq:cploss}. The curves are averaged over four runs with the orange and blue bands indicating the minimum and maximum performance.}
    \label{fig:vbf_cpodd_ndata}
\end{figure}
%--------------------------------

The efficiency of \pysr is also helpful if the amount of training data is reduced. This is shown in
Fig.~\ref{fig:vbf_cpodd_ndata}, where we vary the amount of training data with $\cHWtil = \pm 1$ and show the significance for distinguishing $\cHWtil = 1$ from the SM.
We show two curves for which the observables are learned with and without punishing non-\cpo formulas. 
The learned \pysr formulas offer good performance for as little as 1000 training events. Punishing non-\cpo formulas during the training does affect the performance, but it ensures that the learned formulas are actually \cpo.

\snet requires ten times more events than \pysr to reach a performance plateau. Punishing \cpo formulas decreases the \snet performance significantly, because the algorithm starts with a complicated and not \cpo formula. The \cpo loss then dominates the combined loss, which can only be effectively reduced by deactivating elements of the network early on. These elements, however, can be important for the actual differentiation between the $\cHWtil = \pm 1$ samples, worsening the performance if including the \cpo loss contribution. Finally, the BDT does not reach the data efficiency of \pysr, either.

%%%%%%%%%%%%%%%%%%%%%%%%%%%%%%%%%%%%%%%%%%%%%%%%%%%%%%%
\section{Collin-Soper angle for \texorpdfstring{$t\bar t H$}{ttH} production}
\label{sec:ttH}

Top-associated Higgs production allows to test for \cp violation through the dimension-4 coupling~\cite{Artoisenet:2013puc} 
\begin{align}
    \lag_\text{Yuk} = - \frac{y_t}{\sqrt{2}} \overline{t} \left( c_t + i \gamma_5 \tilde{c}_t \right) t H \; ,
\label{eq:HCM_top}
\end{align}
where $c_t$ and $\tilde{c}_t$ describe the \cp-even and \cp-odd contributions to the coupling, respectively. $t$ is the top-quark field; and $H$, the physical Higgs field.

The corresponding \cp angle is 
\begin{align}
  \tan \alpha_t = \frac{\tilde{c}_t}{c_t} \; ,
\label{eq:def_alpha}
\end{align}
and a value of $45^\circ$ approximately corresponds to the current experimental limits~\cite{ATLAS:2020ior,CMS:2020cga,CMS:2021nnc,CMS:2022dbt}.

This modification of the SM top-Yukawa interaction, which is recovered by setting $c_t = 1$ and $\tilde c_t = 0$, is generated by the dimension-6 SMEFT operator
\begin{align}
    \frac{c_{t\phi}}{\Lambda^2}(\varphi^\dagger\varphi)(\bar Q_3 t_R \widetilde\varphi)\;,
\end{align}
where $Q_3$ is the third-generation quark-doublet and $t_R$ the right-handed top-quark field. The relation between $c_{t\phi}$ and $c_t$/$\tilde c_t$ can be found e.g.~in Ref.~\cite{Fuchs:2020uoc}. Besides the modification of the Higgs--top interaction above, this operator also introduces a Higgs--Higgs--top interaction, which is, however, irrelevant for $t\bar t H$ production.

The top-Yukawa interaction of Eq.\eqref{eq:HCM_top} gives rise to three terms in the squared matrix element, just like in Eq.\eqref{eq:mat_fact},
\begin{align}
    |\mathcal{M}_\text{\ttH}|^2 = c_t^2~|\mathcal{M}_\text{\cp-even}|^2 + 2 c_t \tilde{c}_t~\text{Re} \left[ \mathcal{M}_\text{\cp-even} \mathcal{M}_\text{\cp-odd}^* \right] + \tilde{c}_t^2 |\mathcal{M}_\text{\cp-odd}|^2
\label{eq:top_matrix}
\end{align}
The terms proportional to the squared coupling modifiers are \cp-even, while the interference term is \cp-odd. 

At the LHC, \cp information about the top-Yukawa coupling is primarily obtained from \cp-even, but \cp-sensitive, observables. \cp-odd observables probing the interference term are numerically suppressed~\cite{Azevedo:2022jnd,Barman:2021yfh}. We also show in App.~\ref{sec:ttH_cpodd} that learning optimal \cpo observables does not boost the sensitivity to an observable level.

A powerful \cp-sensitive observable is the Collins-Soper (CS) angle~\cite{Collins:1977iv,Goncalves:2018agy,Barman:2021yfh,Ackerschott:2023nax,Bahl:2024tjy}
\begin{equation}
    \cos \theta^* = \frac{\Vec{p}_t \cdot \Vec{n}}{\Vert \Vec{p}_t \Vert \cdot \Vert \Vec{n} \Vert}
\end{equation}
which is the angle between the \ttbar system and the beam axis $n$ in the \ttbar rest frame. Measuring it requires a reconstruction of the \ttbar system and is therefore challenging. We train symbolic regression models to reconstruct $\theta^*$ from the final-state momenta, so we can assess \cp information without explicitly reconstructing both top quarks.

%%%%%%%%%%%%%%%%%%%%%%%%%%%%%%%%%%%%%%%%%%%%%%%%%%%%%%%%%%%%%%%%%%%
\subsection{Events and training}
\label{subsec:event_gen_tth}

We use \madgraph~3.5.4~\cite{Alwall:2011uj} to separately generate leading order events for
\begin{align}
 pp \to \ttH 
 \qqquad \text{and} \qqquad 
 pp \to \ttH + j 
\end{align}
production. The events are scaled to the NLO rate via a constant $K$-factor of $1.13$ \cite{Demartin:2014fia}. \cp violation in the complex top-Yukawa coupling is introduced via the \texttt{HC\_NLO\_X0 UFO} model~\cite{Artoisenet:2013puc}. The semi-leptonic top decays are simulated with \texttt{MadSpin}~\cite{Artoisenet:2012st}. The events are at parton level, but we will discuss approximate detector effects later. The acceptance cuts are minimal~\cite{Cepeda:2019klc},
\begin{alignat}{9} 
   p_{T,j} > 15~\gev 
   \qquad \text{and} \qquad
   |\eta_{j,\ell}| < 4 \; .
\end{alignat}
The \pysr and \snet training datasets contain the 4-momenta of each final state particle, normalized by the top mass, to obtain dimensionless quantities of order one. \pysr takes individual 4-vector components as input, while \snet takes the entire 4-vector. Since we decay the tops semi-leptonically, we assume that the $b$-jets have been correctly assigned to the lepton and the light quarks.

%--------------------------------
\begin{table}[b!]
\centering
\setlength{\arraycolsep}{5pt}
\renewcommand{\arraystretch}{1.4}
\begin{small}
\begin{equation*}
\begin{array}{ c|cccc } 
\toprule
 \text{Scenario} & \text{Frame} & \nu~\text{Info} & \text{QCD Radiation} & \text{Smearing} \\
 \midrule
 1 & \ttbar & \text{Full} & \times & \times \\
 2 & \text{Lab} & \text{Full} & \times & \times \\
 3 & \text{Lab} & E_T^{\text{miss}} & \times & \times \\
 4 & \text{Lab} & E_T^{\text{miss}} & \checkmark & \times \\
 5 & \text{Lab} & E_T^{\text{miss}} & \times & \checkmark \\
 6 & \text{Lab} & E_T^{\text{miss}} & \checkmark & \checkmark \\
 \bottomrule
\end{array}
\end{equation*}
\end{small}
\caption{Benchmark scenarios to assess the performance of \pysr and \snet. The scenarios become increasingly more complex.}
\label{tab:scenarios}
\end{table}
%--------------------------------

To assess the performance in reconstructing $\theta^*$, we define six benchmark scenarios for the information given to \pysr and \snet, listed in Tab.~\ref{tab:scenarios}. The first two scenarios include the full kinematic information at parton level, but in different rest frames. From scenario~3 on, the full neutrino momentum is replaced by $E_T^\text{miss}$. Scenarios~4 and 6 include an additional hard jet, while scenarios 5 and 6 approximate limited detector resolution via smearing. For scenarios~1 and 2, where the full kinematic information is available, we use an MSE loss as $\mathcal{L}_\text{base}$. Otherwise, we use an inverse Gaussian loss
\begin{align}
    \loss_\text{InvGaussian} = 1 - \exp \left[ -\frac{(y - \hat{y})^2}{2 \left( \dfrac{\max(y)}{\sigma} - \dfrac{\min(y)}{\sigma} \right)^2 }\right] \;.
\end{align}
with $\sigma = 8$, which is more robust against outliers. Details on this loss function are provided in \cref{app:loss}.

For the training of \pysr, the hyperparameters are slightly varied depending on the scenario. In scenario 1, we train for 200 iterations and allow a maximum complexity of 60. The functions given to \pysr are the square, square root, addition, and division. For the other scenarios, we expand the functions to also include the sine, cosine, sine hyperbolicus, subtraction, and multiplication. Furthermore, the training iterations are raised to 2000 (3000) and the maxsize to 50 (60) in scenarios~2 to 4 (5 and 6). 

Ensuring a stable training of \snet is non-trivial, because the activation functions are replaced with various mathematical operations. Furthermore, the sparsity training can quickly lead to gradient instabilities due to the sudden jumps in the loss when a threshold is reached, pruning a weight or an operator. To improve the training, we split it into the three stages introduced in Sec.~\ref{subsec:symbolnet}: First, we train with disabled sparsity thresholds, corresponding to a default MLP training. We found that a small initial weight initialization with a uniform distribution over $[-0.02, 0.02]$ gives the most stable results. 

%--------------------------------
\begin{table}[t]
\centering
\setlength{\arraycolsep}{5pt}
\renewcommand{\arraystretch}{1.4}
\begin{small}
\begin{equation*}
\begin{array}{ c|c|c|c|c|c|c|c } 
\toprule
 \text{Scenario} & N_{\text{epochs}}^{\text{default}} & N_{\text{epochs}}^{\text{mixed}} & N_{\text{epochs}}^{\text{sparsity}} & LR^{\text{default}} & LR^{\text{mixed}} & LR^{\text{sparsity}} & BS \\
 \midrule
 1 & 200 & 100 & 20 & 0.01 & 0.005 & 0.001 & 128 \\
 \midrule
 2 & 500 & 100 & 20 & 0.01 & 0.005 & 0.001 & 128 \\
 \midrule
 3~\&~4 & 1000 & 400 & 50 & 0.01 & 0.002 & 0.001 & 64 \\
 \midrule
 5~\&~6 & 2000 & 500 & 50 & 0.02 & 0.002 & 0.001 & 64 \\
 \bottomrule
\end{array}
\end{equation*}
\end{small}
\caption{Hyperparameters for the training of \snet in the various scenarios.}
\label{tab:hyperparams}
\end{table}
%--------------------------------

%--------------------------------
\begin{table}[b!]
\centering
\setlength{\arraycolsep}{5pt}
\renewcommand{\arraystretch}{1.4}
\begin{small}
\begin{equation*}
\begin{array}{ c|c|c|c|c|c } 
\toprule
 \text{Scenario} & \text{Operators} & V \to V~\text{layer} & V \to S~\text{layer} & S \to S~\text{layer} & S \to S~\text{layer} \\
 \midrule
 \multirow{2}{*}{1} & f(p) & - & \{p_z, \Vert p \Vert_3\} & \{id.\} & - \\
 & g(p_i, p_j) & - & \{ \langle p_i \times p_j \rangle_3 \} & \{ / \} & - \\
 \midrule
 \multirow{2}{*}{2} & f(p) & \{ id. \} & \{p_z, \Vert p \Vert_3\} & \{id.\} & - \\
 & g(p_i, p_j) & \{ \mathrm{boost}\} & \{ \langle p_i \times p_j \rangle_3 \} & \{ / \} & - \\
 \midrule
 \multirow{2}{*}{3~\&~4} & f(p) & \{ \tanh \} & \{p_z, \Vert p \Vert_3\} & \{~^2, \sqrt, \sin, \cos\} & - \\
 & g(p_i, p_j) & \{ \mathrm{boost}\} & \{ \langle p_i \times p_j \rangle_3 \} & \{ *, / \} & - \\
 \midrule
 \multirow{2}{*}{5~\&~6} & f(p) & \{ \tanh \} & \{p_0, p_x, p_y, p_z, \Vert p \Vert_3\} & \{~^2, \sqrt\} & \{\sin, \cos\} \\
 & g(p_i, p_j) & \{ \mathrm{boost}\} & \{ \langle p_i \times p_j \rangle_3, \langle p_i \times p_j \rangle_4 \} & \{ *, / \} & \{ +, - \} \\
 \bottomrule
\end{array}
\end{equation*}
\end{small}
\caption{Types of layers and functions used for the training of \snet in the various scenarios.}
\label{tab:functions}
\end{table}
%--------------------------------

After this training, the network with the lowest validation loss is passed to the mixed training, where weights and sparsity thresholds are both trainable parameters. Finally, the weights are disabled and the sparsity thresholds are trained alone. During the latter two training phases, the loss and gradient are monitored, and training is stopped when one of them diverges. We perform hyperparameter scans for each scenario, varying the learning rate, batch size and model complexity. The hyperparameters can be found in \cref{tab:hyperparams}. For optimization, the \texttt{Adam} optimizer was used. The \texttt{LookaheadAdam} optimizer was considered, but did not lead to improved results. The functions used in each layer can be found in \cref{tab:functions}.

%%%%%%%%%%%%%%%%%%%%%%%%%%%%%%%%%%%%%%%%%%%%%%%%%%%%%%%%%%%%%%%%%%%
\subsection{Results}
\label{subsec:results_tth}

The \pysr and \snet predictions for the six scenarios are depicted in Fig.~\ref{fig:CS_full_perfect}. In the simplest scenario~1, the SR algorithms only need to build the top quarks from the provided decay products and learn the angle with respect to the beam axis. Both algorithms perform well and find the exact structure of the analytic formula,
\begin{align}
\cos \theta^*_{\pysr} &= \frac{p_{z,b} + p_{z,\bar{l}} + p_{z,\nu}}{\sqrt{\left(p_{x,b} + p_{x,\bar{l}} + p_{x,\nu}\right)^{2} + \left(p_{y,b} + p_{y,\bar{l}} + p_{y,\nu}\right)^{2} + \left(p_{z,b} + p_{z,\bar{l}} + p_{z,\nu}\right)^{2}}}\;, \notag \\[2ex]
\cos \theta^*_{\snet} &= \frac{1.006 p_{z,b} + 1.001 p_{z,\bar{l}} + 1.002 p_{z,\nu} - 1.027 p_{z,\bar{b}} - 1.027 p_{z,q} - 1.031 p_{z,\bar{q}}}{\Big\Vert 1.034 p_{b} + 1.022 p_{\bar{l}} + 1.024 p_{\nu} - p_{\bar{b}} - 1.007 p_{q} - 1.009 p_{\bar{q}}\Big\Vert_3} \notag \\[2ex]
&\approx \frac{1}{2} \bigg( \frac{ p_{z,t} }{ \Vert p_t \Vert_3 } - \frac{ p_{z,\bar{t}} }{ \Vert p_{\bar{t}} \Vert_3 } \bigg) \;,
\end{align}
where all variables are defined in the \ttbar frame. The only difference is that \snet does not tune all weights to exactly one. We traced this back to numerical noise in \textsc{MadSpin}. Moreover, \snet does not exclude one side of the top decay. Instead, it builds the CS angle twice --- once from the leptonic and once from the hadronic top decay. The resulting numerical differences to the true formula are negligible. For the other scenarios, the learned formulas are given in \cref{app:formulas}. In scenario~2 the variables are given in the lab frame. \snet reaches a similar accuracy as before, since it can apply a boost to the input variables before building the formula. In contrast, \pysr cannot mimic a boost from the scalar input variables and shows slight deviations from the truth.

%--------------------------------
\begin{figure}
    \includegraphics[width=.49\textwidth]{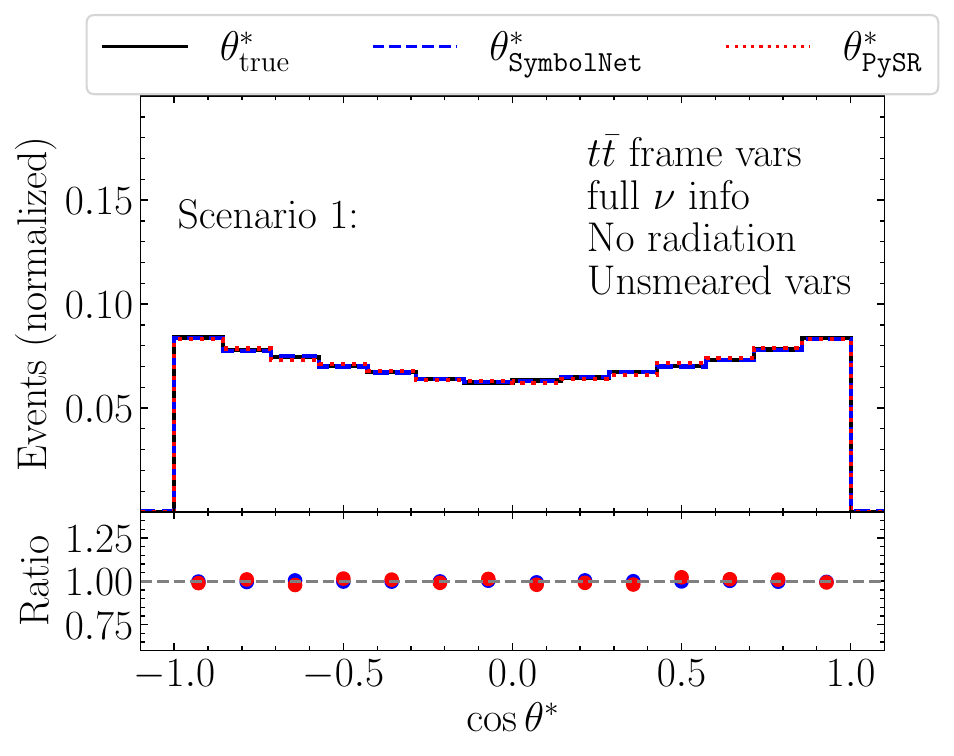}
    \includegraphics[width=.49\textwidth]{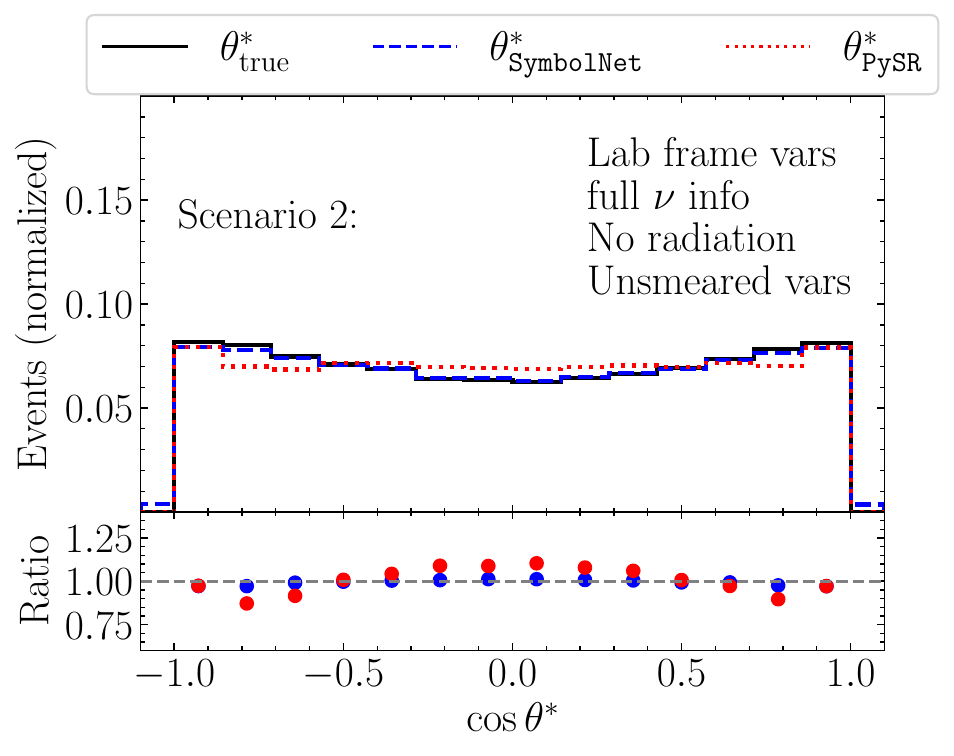} \\
    \includegraphics[width=.49\textwidth]{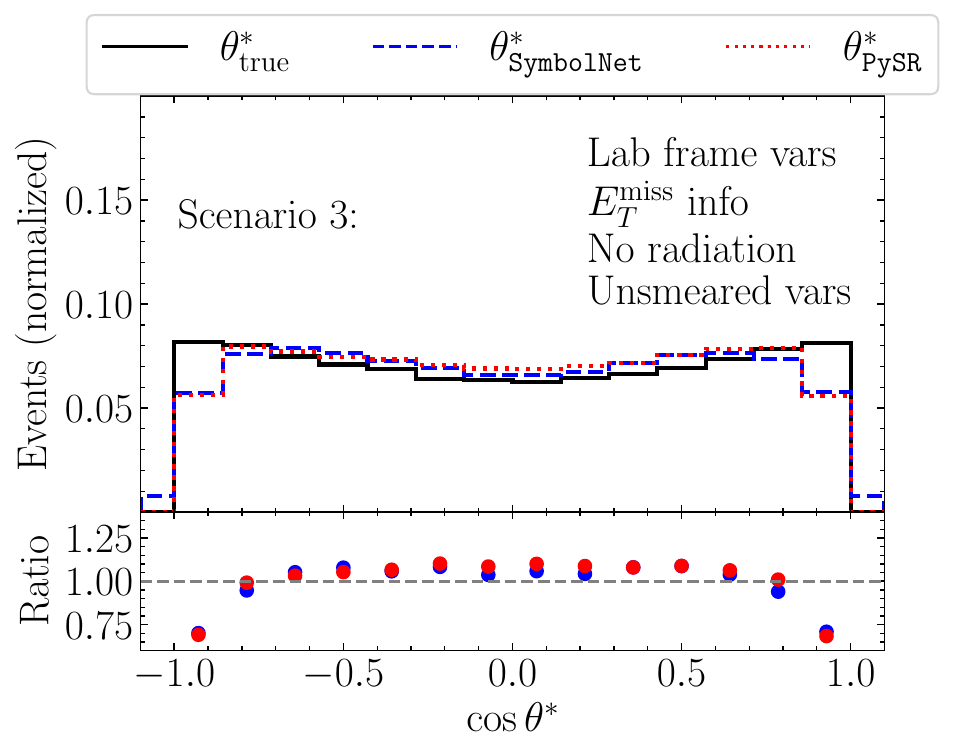}
    \includegraphics[width=.49\textwidth]{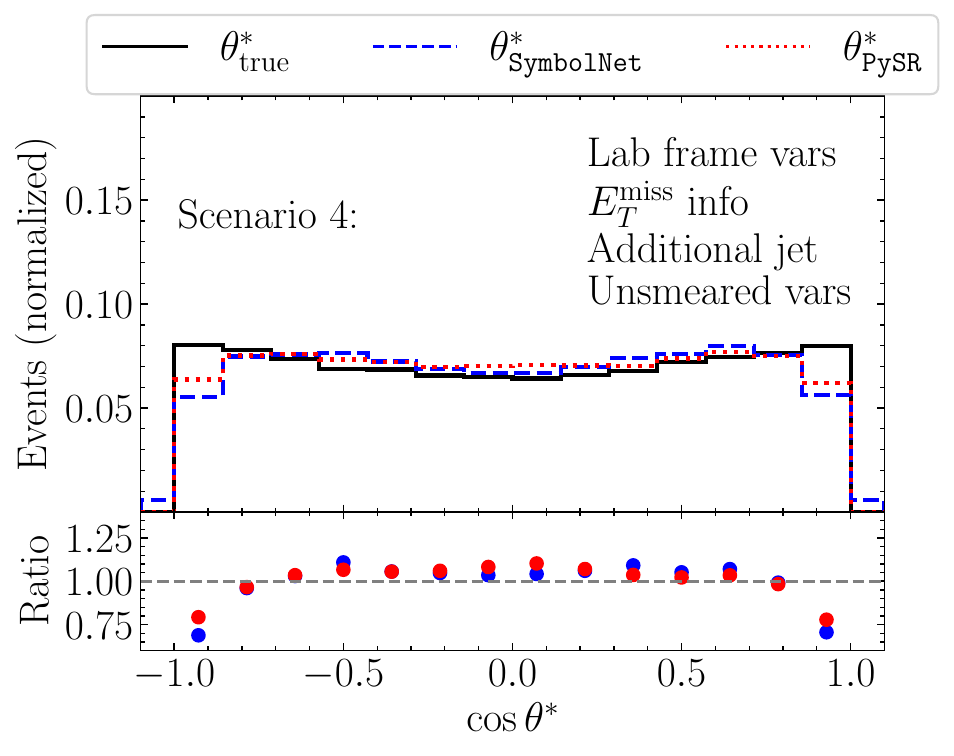} \\
    \includegraphics[width=.49\textwidth]{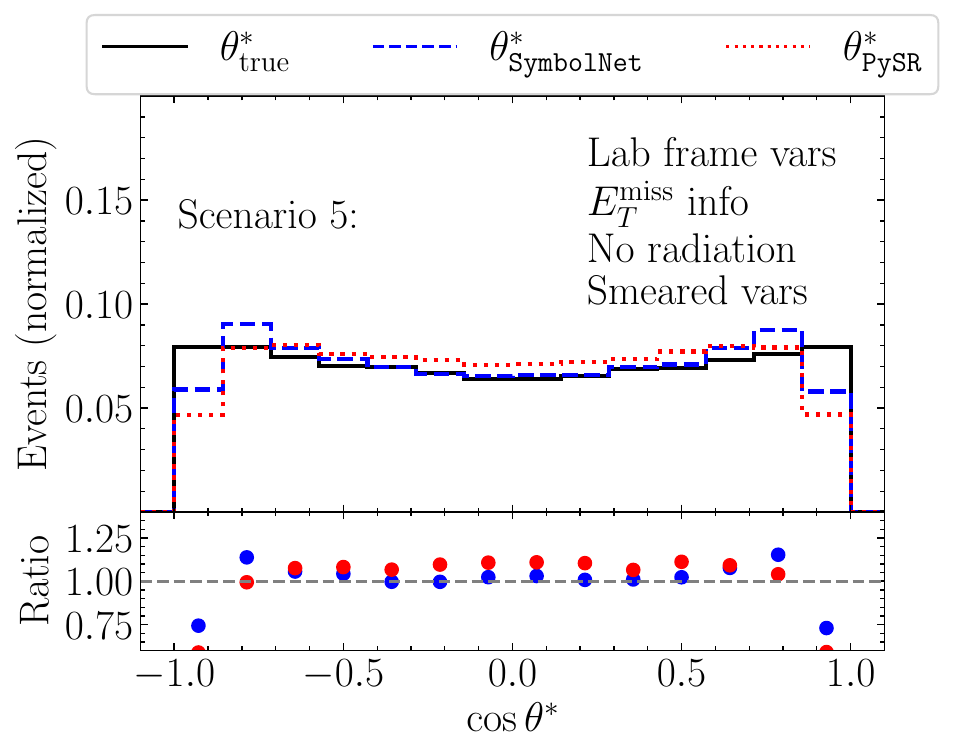}
    \includegraphics[width=.49\textwidth]{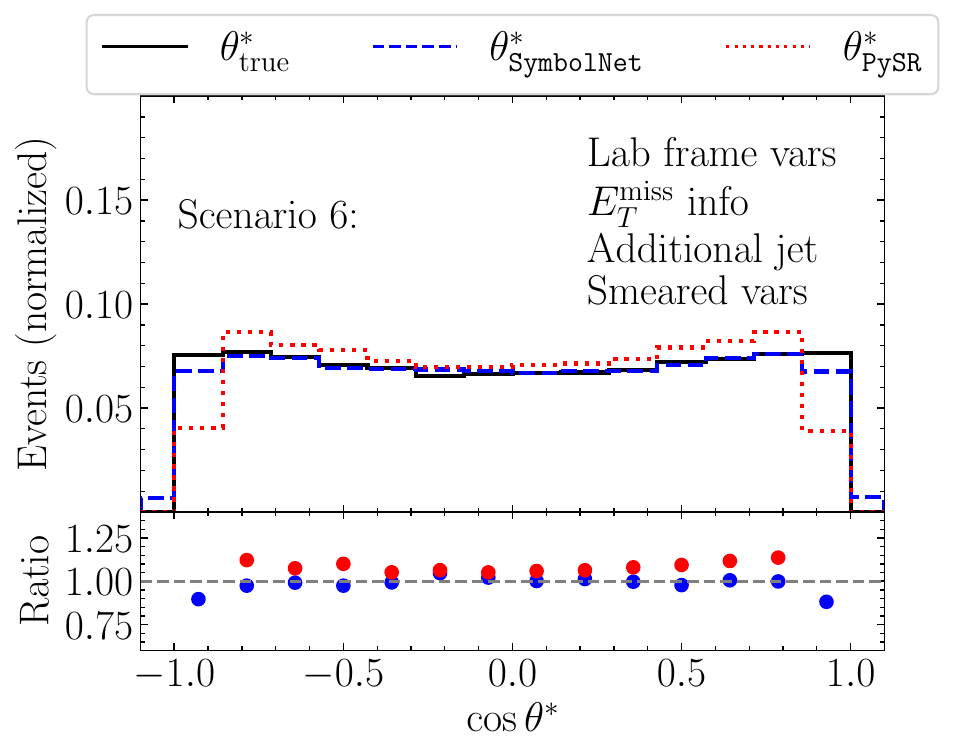}
    \caption{Predicted distributions of the CS angle for the six scenarios defined in Tab.~\ref{tab:scenarios}. Two overflow bins are added to each distribution.}
    \label{fig:CS_full_perfect}
\end{figure}
%--------------------------------

In all other scenarios the missing neutrino information implies that there is no analytic reference formula. Also, since the reconstruction tasks get more and more difficult, the goodness of the fit decreases with increasingly realistic scenarios. In scenarios~3 and 4, where no smearing is applied, \pysr and \snet yield very similar results. When the smearing is added for the results in the lower row of Fig.~\ref{fig:CS_full_perfect}, \snet performs better than the simple \pysr formula
\begin{align}
\cos \theta^*_{\pysr} =
& \sin \bigg[\bigg( 1.114 p_{z,b} + 2.143 p_{z,\bar{l}} - 0.858 p_{z,\bar{b}} - 0.426 p_{z,q} - 1.088 p_{z,\bar{q}} \bigg) \bigg/ \notag \\
& \bigg( E_T^{\text{miss}} E_{q} + E_{b} + E_{\bar{b}} + 1.85 E_{\bar{l}} + E_{\bar{q}} \notag \\
& - \left(p_{x,b} + p_{x,\bar{l}}\right) \left(p_{x,\bar{b}} + p_{x,q} + p_{x,\bar{q}}\right) - 0.863 - \frac{0.205 p_{z,q}}{\sqrt{E_{q}}} \bigg) \bigg]
    \; .
    \label{eq:thetaPySR56}
\end{align}
It yields more accurate predictions for all values of $|\cos \theta^*|$ and consequently also has a lower MSE loss, as detailed below.

An interesting observation in Fig.~\ref{fig:CS_full_perfect} is that both \snet and \pysr accurately fit the regime around $\cos \theta^* \approx 0$, but underestimate the extreme bins with $|\theta^*| \approx 1$. This holds true whenever there is no true analytic formula to be found. A possible explanation is offered by the cyclic property of the CS angle. Any angle outside $\cos \theta^* \in [-1, 1]$ is automatically mapped back into the interval, since it corresponds to the same physical state. This behavior has to be learned via restricting the output to this range, which does not always happen exactly. In some scenarios, \snet predicts values outside of this range, as indicated by the overflow bins in \cref{fig:CS_full_perfect}. For \pysr, the final output functions are often restricted to a smaller range, e.g. $\cos \theta^* \in [-0.95, 0.95]$, which might explain the underestimation in the outer bins. We found that neither a cyclic loss nor a fixed mapping of the output layer to the range $[-1, 1]$ improve the results. 

\paragraph{Structure of learned formulas} Although the final formulas shown above and in \cref{app:formulas} are not very intuitive, there exist some general patterns. First, \pysr formulas are much less complex, as it starts from a very simple formula and then increases the complexity. Despite having to predict $\cos \theta^*$, \pysr always chooses the core of the formulas as
\begin{align}
    \cos \theta^*_{\pysr} \sim \sin \frac{\sum_i a_i~p_{i,z}}{\sum_i b_i~E_i} \; .
\label{eq:tth_pysr}
\end{align}
For low complexity, \pysr usually uses the energy as a replacement for the 3-vector norm. Similarly, the \snet formulas have a universal structure, but it is more complex,
\begin{align}
    \cos \theta^*_{\snet} \sim  \frac{\operatorname{boost} \left( \sum_i a_i~p_{i}~|~\sum_j b_j~p_{j} \right)\bigg|_{z}}{\bigg \Vert \operatorname{boost} \left( \sum_i a_i~p_{i}~|~\sum_j b_j~p_{j} \right) \bigg\Vert_3}  \; \; \; .
\label{eq:tth_snet}
\end{align}
This formula reduces to the exact CS angle after including the full neutrino information and for $a_i = b_i = 1$. The two general forms indicate that the structure of the parton-level formulas is also useful at the reco level. 

%--------------------------------
\begin{figure}[t]
    \centering
    \includegraphics[width=.6\textwidth]{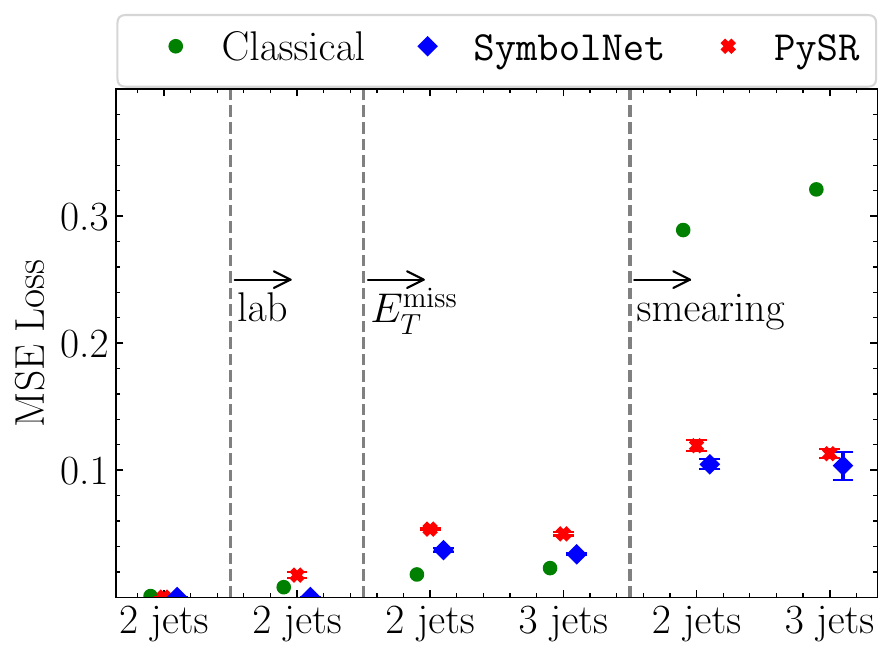}
    \caption{MSE values for each scenario. The values are shown for the classical reconstruction (green), \pysr (red), and vectorized \snet (blue). The central values and error bars correspond to the mean and the standard deviation of ten independent training runs.}
    \label{fig:mse_etacut}
\end{figure}
%--------------------------------

In Fig.~\ref{fig:mse_etacut}, we show the MSE values for the \pysr (red) and \snet (blue) formulas evaluated on the test dataset. For each of the six benchmarks both algorithms are trained ten times, and we show the mean and standard deviation of the computed MSE. Runs that do not converge are discarded. The MSE is not necessarily the training objective, instead we use it to assess the goodness of the fit. The MSE values of the two formulas are contrasted with a classical reconstruction of the top quarks~\cite{Barman:2021yfh}, where the longitudinal neutrino momentum is reconstructed using the $W$-mass constraint. After that, the light jets, $b$-jets and $W$-boson are combined to two top quarks. We see that this classical reconstruction leads to slightly better MSE values when the neutrino information is missing, but the variables are not smeared. Including detector smearing, the classical reconstruction is clearly outperformed by \pysr and by \snet. Among the two learned formulas, \snet leads to slightly lower MSE values than \pysr. 

%--------------------------------
\begin{figure}
    \centering
    \includegraphics[width=.49\textwidth]{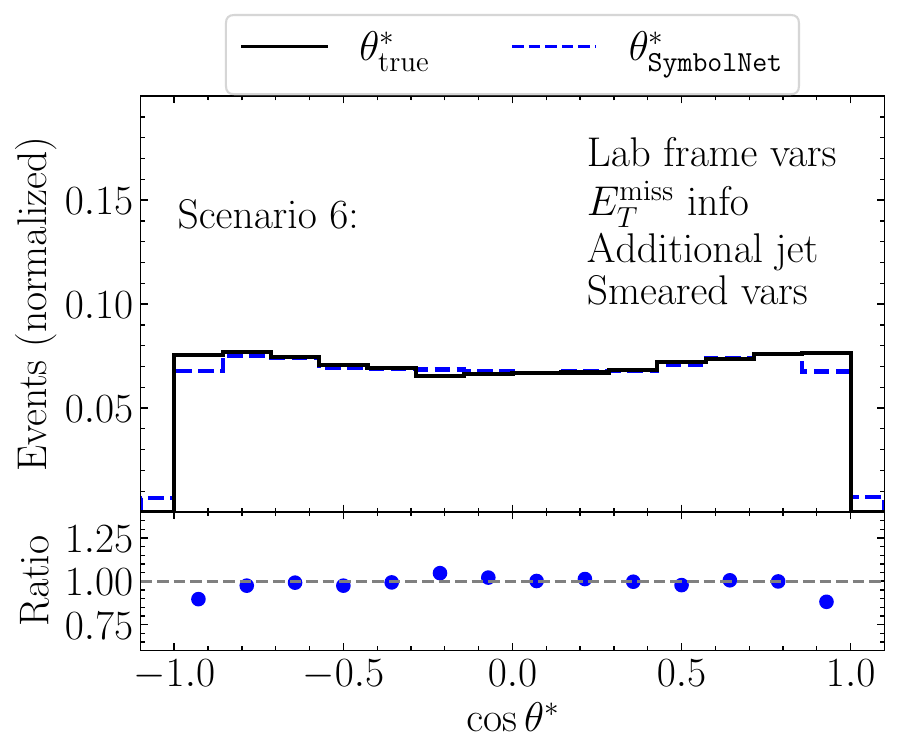} 
    \includegraphics[width=.49\textwidth]{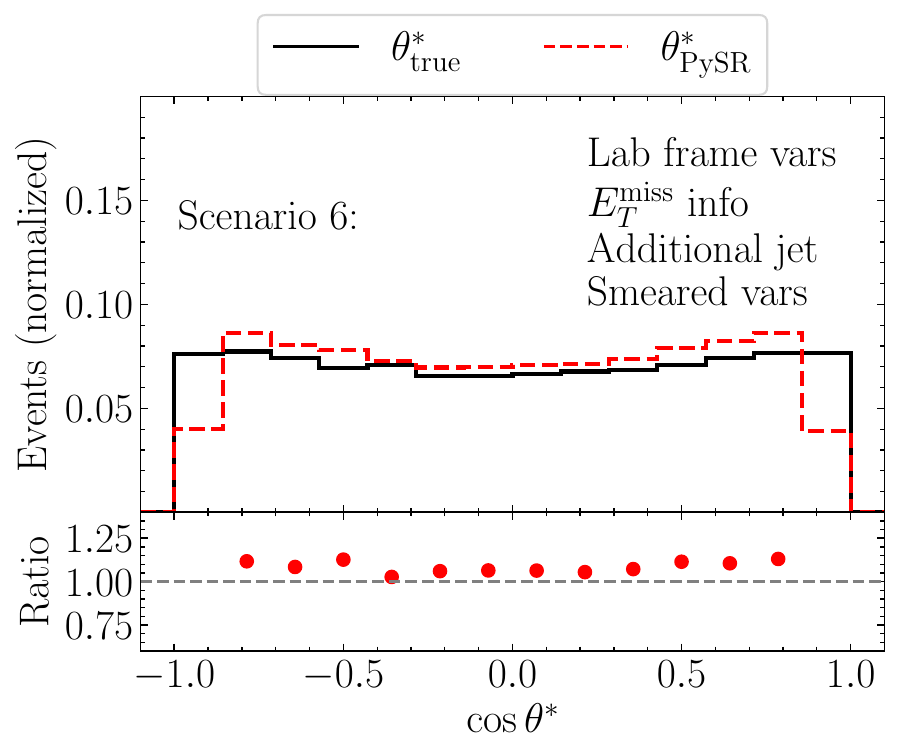} \\
    \includegraphics[width=.435\textwidth]{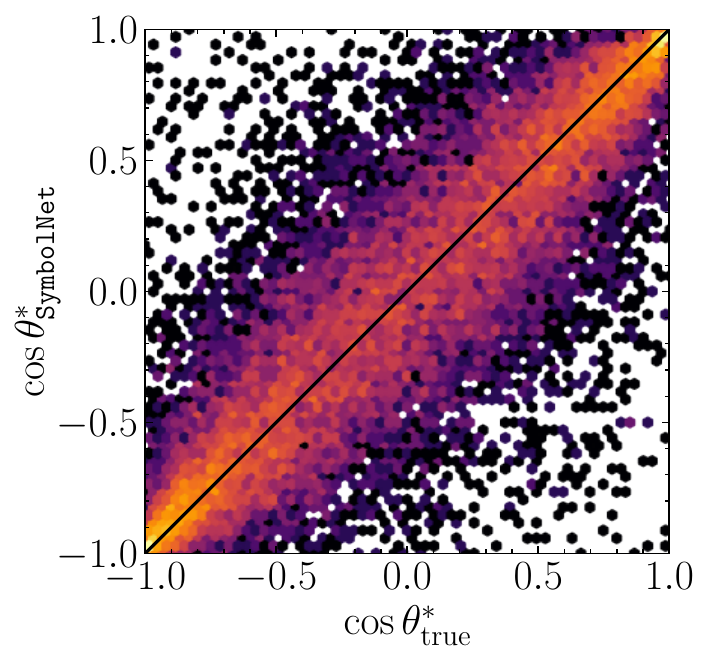}
    \includegraphics[width=.435\textwidth]{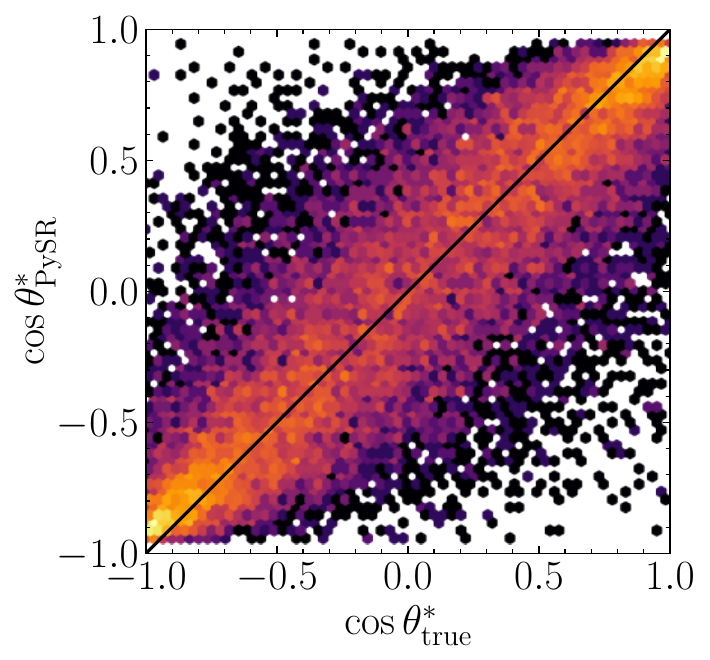}
    \includegraphics[width=.08\textwidth]{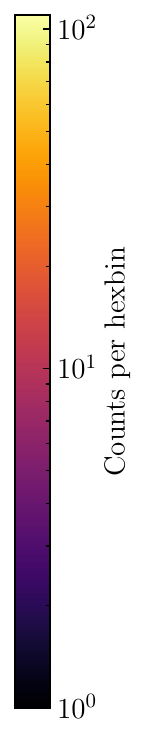} \\
    \includegraphics[width=.49\textwidth]{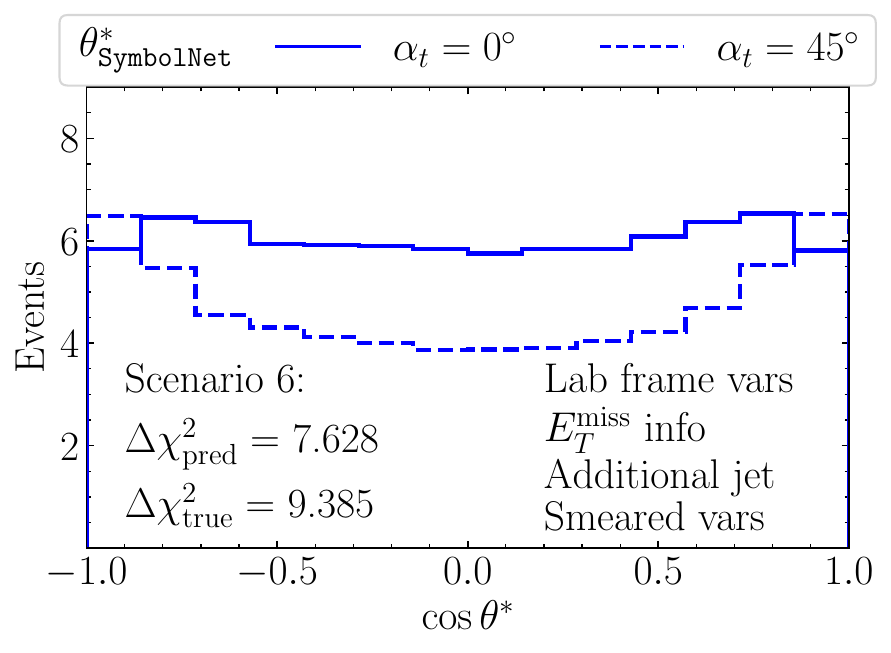}
    \includegraphics[width=.49\textwidth]{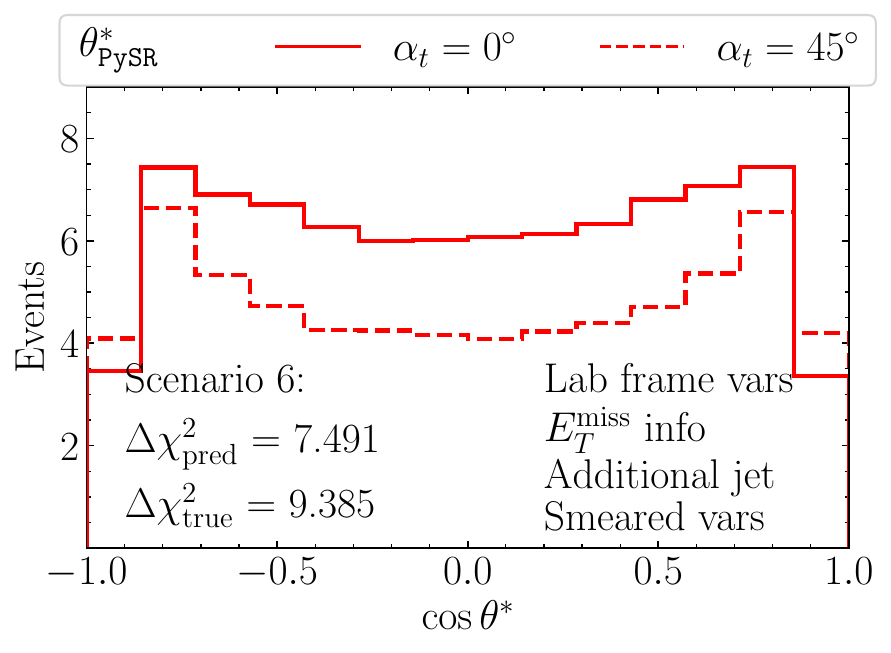}
    \caption{\snet (left) and \pysr (right) formulas for scenario 6 with perfect $b$-ordering. We show the
    learned CS angle, its calibration normalized to the same maximum count of events in one bin, and the distributions for $\alpha_t = 45^\circ$ including the expected \cp-sensitivity.
    }
    \label{fig:CS_full_perfect_6}
\end{figure}
%--------------------------------

\paragraph{Observable performance} 
In \cref{fig:CS_full_perfect_6}, we compare \pysr and \snet for scenario~6, with QCD radiation and detector smearing. The upper panels show the distribution of $\cos \theta^*$ learned by \snet and \pysr analytically. \snet slightly underestimates the extreme bins and shifts the missing events in these bins outside of $[-1,1]$. This way the central regime of the distribution is accurately reproduced.
\pysr, with a formula restricted to $[-0.93, 0.93]$, dramatically fails in the extreme bins, leading to all central bins coming out high. Below, we show the correlation of learned and true CS-angles for all training events. For \snet, the events closely follow the diagonal line, while the \pysr formula leads to a small tilt. The bins with the highest number of events are at large $|\cos \theta^*|$, but for \pysr they are shifted away from the diagonal, resulting in the observed underestimation of the extreme bins. The width of the colored area is smaller for \snet, indicating a more accurate formula. \snet also has more outliers away from the diagonal, but these single events can be attributed to statistical fluctuations.

In the lower panels in Fig.~\ref{fig:CS_full_perfect_6}, we compare the \cp sensitivity of the two learned formulas, testing $\alpha_t = 45^\circ$ against the SM hypothesis $\alpha_t = 0^\circ$. For this hypothesis test, the two formulas are trained on SM events. The number of expected events are taken from Ref.~\cite{Bahl:2024tjy}. They are based on ATLAS analyses of the \ttH channel and assume a luminosity of $\mathcal{L} = 300\ifb$. For $\alpha_t = 45^\circ$ the rate is smaller in the SM. Both learned formulas capture the main feature of the $45^\circ$ case, namely that the CS distribution develops clear maxima for the extreme bins. Distinguishing the two datasets, \snet reaches a sensitivity of $\Delta\chi^2_{\snet} = 7.628$, compared to $\Delta\chi^2_{\pysr} = 7.491$ and the parton level sensitivity $\Delta\chi^2_\text{true} = 9.385$.

%--------------------------------
\begin{figure}[t]
    \centering
    \includegraphics[width=.6\textwidth]{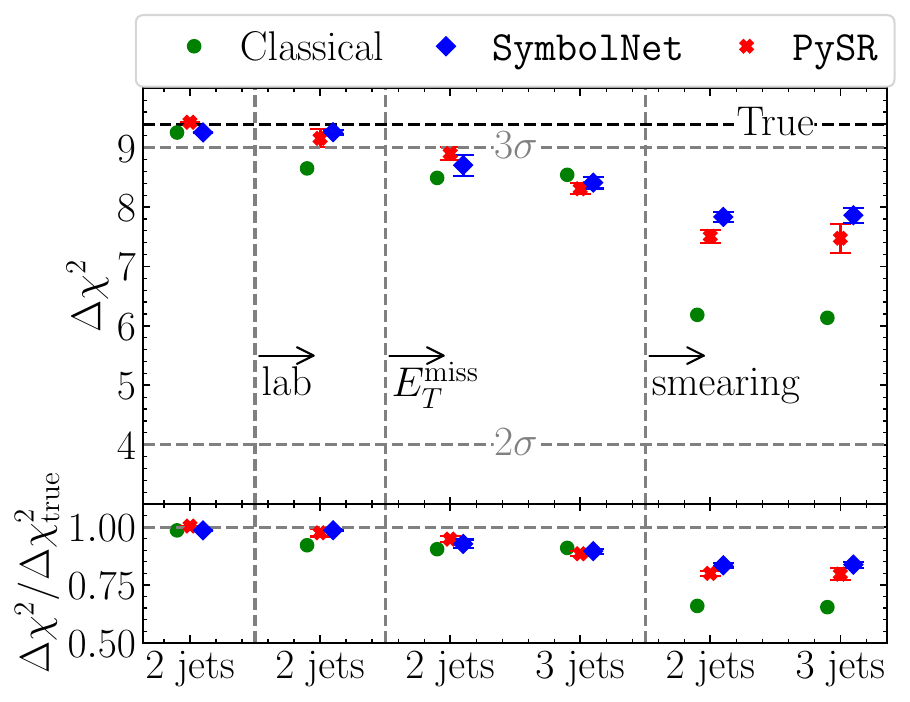}
    \caption{$\Delta \chi^2$ values for excluding $\alpha_t = 45^\circ$ based on a measured SM-dataset. We show results for a classical reconstruction (green),  \pysr (red), and vectorized \snet (blue), compared to the parton-level $\chi^2$. The central values and error bars correspond to the mean and the standard deviation of ten independent trainings.}
    \label{fig:chi2_etacut}
\end{figure}
%--------------------------------

%--------------------------------
\begin{figure}[t]
    \centering
    \includegraphics[width=.6\textwidth]{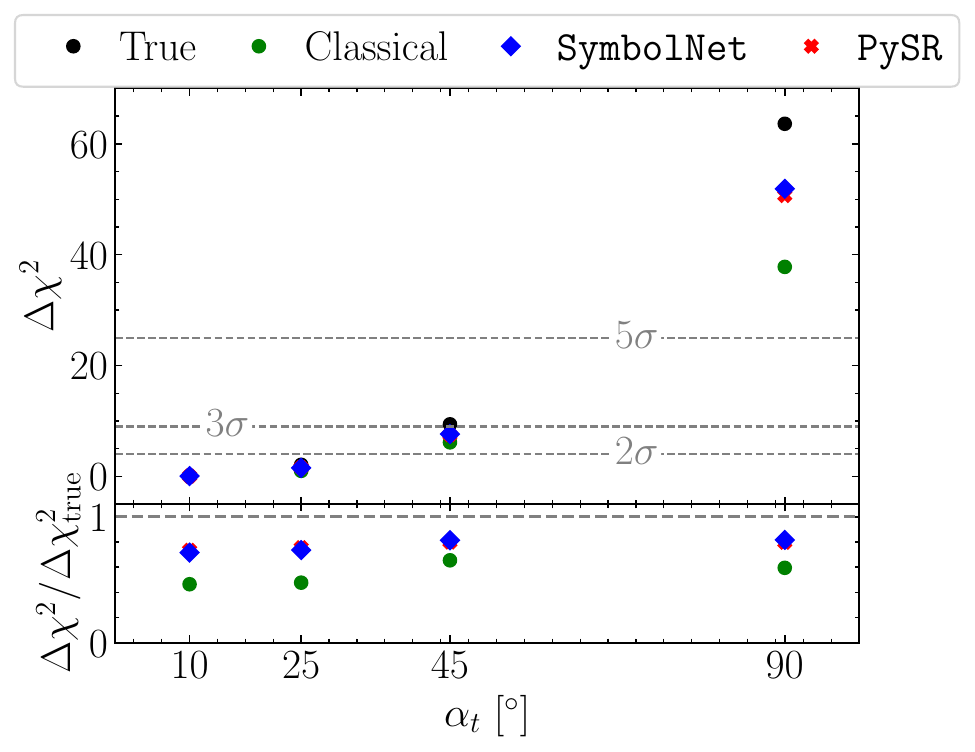}
    \caption{$\Delta \chi^2$ values from the \pysr and \snet formulas for scenario 6 by comparing the SM $\alpha_t = 0^\circ$ hypothesis with different amounts of \cp violation.
    }
    \label{fig:chi2_compare}
\end{figure}
%--------------------------------

Fig.~\ref{fig:chi2_etacut} shows the $\Delta \chi^2$ values in each scenario for \snet and \pysr and compares them to the classical reconstruction algorithm. In scenario~1, all three methods get very close to the true parton-level information. The slightly lower results are an effect of \texttt{MadSpin} introducing small fluctuations in the decayed particle momenta. For the classical reconstruction, this effect becomes more drastic when the reconstructed top quarks are boosted. On the other hand, \pysr and \snet manage again to yield $\Delta \chi^2$ values close to the parton level case. In scenarios~3 and~4, the \cp information drops slightly for all methods, but still yielding similar results. The picture changes with detector smearing, where the classical reconstruction algorithm only captures around $60\%$ of the total \cp information. \pysr and \snet restore up to $80\%$ of the \cp information in these realistic scenarios, with \snet giving slightly better results. 

Finally, we investigate the performance of our methods for different \cp angles in Fig.~\ref{fig:chi2_compare}. We show $\Delta \chi^2$ from \pysr, \snet, and the classical reconstruction algorithms in scenario~6 for $\alpha_t = [10^\circ, 25^\circ, 45^\circ, 90^\circ]$. The results for $\alpha_t = 45^\circ$ correspond to our above discussion. For the fully \cp-odd choice $\alpha_t = 90^\circ$, the same relative \cp information of \ts is extracted by all three methods. For smaller angles, the performance of all methods drops, because of the less pronounced differences to the SM-kinematics and the greater impact of the smearing. Still, \snet and \pysr show a clear advantage over the classical reconstruction algorithm.

%%%%%%%%%%%%%%%%%%%%%%%%%%%%%%%%%%%%%%%%%%%%%%%%%%%%%%%
\section{Conclusions}
\label{sec:conclusions}

Searching for \cp violation in the Higgs sector is a well-motivated but challenging task. Modern ML has been applied with great success to improve these searches. Typical numerical approaches, however, lack the interpretability and control needed to test a fundamental symmetry like \cp. We have shown how interpretable and controlled analytical expressions for testing the \cp nature of the Higgs boson can be obtained using symbolic regression. We employed two complementary SR approaches: \pysr based on a genetic algorithm, starting with a simple expression which is evolved into a more complicated expression; and \snet based on a neural-network-like approach, starting with a complicated expression which is then successively reduced to a simpler expression.

In the first part of the paper, we have shown how to derive optimal \cpo observables at the detector level in analytic form. Focusing on VBF Higgs production, we compared the \cp sensitivity of the observables learned using \pysr and \snet to a numerical BDT approach. We find the observables to outperform the BDT. We also showed that the SR approaches --- in particular \pysr --- are more data efficient, meaning that they are able to analyze data with a very small \cpo component. Investigating the learned analytical expressions, we were able to confirm their \cpo nature explicitly. As expected, the learned \cpo observables have a similar structure as the known parton-level optimal observable. Our learned analytical optimal observables are straightforward to incorporate into actual analyses.

In the second part of the paper, we focused on \ttH production, where \cpo observables are very hard to measure at the LHC. We discussed the reconstruction of the \cp-sensitive parton-level Collins-Soper angle from reco-level data from semileptonic top decays. We studied six scenarios with different levels of available information, such as the neutrino momentum and limited detector resolution. In the most realistic scenarios, we found the SR approaches to be able to reconstruct the parton-level CS angle significantly better than traditional top-reconstruction algorithms. We also showed that the reconstructed CS angle preserves the \cp sensitivity of the true parton-level CS angle.  Due to the better expressivity, \snet performs slightly better than \pysr. Looking at the learned analytic expressions, we were able to identify structures similar to the parton-level CS angle. As for the optimal \cpo observables, the learned analytic approximations of the parton-level CS angle can be directly incorporated into the analysis of real data.

By comparing the \pysr and \snet approaches, we found that \pysr performs better for problems with a small amount of relevant data. If a large amount of data is available and more accuracy is needed, we showed that \snet outperforms \pysr. This demonstrates the complementarity of both approaches. This means that using SR for Higgs \cp analyses not only strengthens the link between fundamental theory and complex experimental analyses, but it can also offer performance advantages in particular if data efficiency is important.

%%%%%%%%%%%%%%%%%%%%%%%%%%%%%%%%%%%%%%%%%%%%%%%%%%%%%%%%%%%%%%%%%%%
%%%%%%%%%%%%%%%%%%%%%%%%%%%%%%%%%%%%%%%%%%%%%%%%%%%%%%%%%%%%%%%%%%%

\section*{Acknowledgements}

HB and TP acknowledge support through the KISS consortium (05D2022) funded by the German Federal Ministry of Education and Research BMBF in the ErUM-Data action plan, by the Deutsche Forschungsgemeinschaft (DFG, German Research Foundation) under grant 396021762 – TRR 257: Particle Physics Phenomenology after the Higgs Discovery, and through Germany’s Excellence Strategy EXC 2181/1 – 390900948 (the Heidelberg STRUCTURES Excellence Cluster) and by the state of Baden-Württemberg through bwHPC and the German Research Foundation (DFG) through grant INST 35/1597-1 FUGG.
EF and MM were funded by the Deutsche Forschungsgemeinschaft (DFG, German Research Foundation) under Germany’s Excellence Strategy – EXC-2123 QuantumFrontiers – 390837967.
The authors acknowledge resources provided by the LUIS computing cluster at Leibniz University Hannover, which is funded by the Deutsche Forschungsgemeinschaft (DFG, German Research Foundation) – Projektnummern INST 187/742-1 FUGG, INST 187/592-1 FUGG, and INST 187/430-1. 
This work has been partially funded
by the Deutsche Forschungsgemeinschaft (DFG, German
Research Foundation) - 491245950.

\appendix

\newpage
%%%%%%%%%%%%%%%%%%%%%%%%%%%%%%%%%%%%%%%%%%%%%%%%%%%%%%%%%%%%%%%%%%%
%%%%%%%%%%%%%%%%%%%%%%%%%%%%%%%%%%%%%%%%%%%%%%%%%%%%%%%%%%%%%%%%%%%

\section{Building \cp-odd observables in \texorpdfstring{$t\bar tH$}{ttH}}
\label{sec:ttH_cpodd}

Dedicated \cp-odd measurements in \ttH are not feasible for current studies at the LHC due to the smallness of the interference term. Despite, or precisely because of this, it is still interesting to test how much sensitivity an optimal \cpo observable can reach, as this is the only way of unambiguously testing for \cp violation. The discussion in \cref{sec:training_cpodd_observable} still holds for \ttH, however, we use a slightly different SR algorithm. 

\cp-odd observables can generally be defined via 
\begin{align}
\label{eq:tth_cpodd_epsilon}
    \epsilon_{\mu\nu\rho\sigma} p_1^\mu p_2^\nu p_3^\rho p_4^\sigma
\end{align}
where the $p_i$ are linearly independent momenta or polarization vectors of the final or initial state particles. When boosted to the rest frame of one of the momenta, the $\epsilon$-tensors reduce to triple products (TP), which are commonly used in analyses targeting the interference term in \ttH production (see e.g.~\cite{Bevan:2014nva,Mileo:2016mxg,Ellis:2013yxa,Boudjema:2015nda,Miralles:2024huv}). For example, in the \ttbar rest frame this results in~\cite{Goncalves:2018agy,Goncalves:2021dcu}
\begin{align}
\label{eq:tth_cpodd_dphi}
    \Delta \phi_{l\bar{l}}^{\ttbar} = \mathrm{sgn} \left[ \Vec{p_t} (\Vec{p_{l^+}} \times \Vec{p_{l^-}}) \right]~\arccos~\left[ \frac{\Vec{p_t} \times \Vec{p_{l^+}}}{|\Vec{p_t} \times \Vec{p_{l^+}}|} \cdot \frac{\Vec{p_t} \times \Vec{p_{l^-}}}{|\Vec{p_t} \times \Vec{p_{l^-}}|} \right] \; .
\end{align}
In particular, in the fully leptonic \ttH decay, a set of $22$ variables, dubbed $\omega_i$, can be defined out of combinations of the final state momenta~\cite{Faroughy:2019ird,Bortolato:2020zcg}. From here on, we will refer to the $\epsilon$-tensors as TPs.

Here, we work in the fully leptonic decay channel but do not fix the momenta in \cref{eq:tth_cpodd_epsilon}. Instead, we let \snet learn \cpo observables by construction using TPs:
\begin{itemize}
    \item We define two instances of \snet which do not share connections but inherit their input from the same input layer.
    \item The first instance of \snet only contains \cp-even variables and is therefore \cp-even by construction. We call the output of this instance $D_\mathrm{even}(x)$. 
    \item The second instance of \snet has a special $V \to S$ layer, which only includes a TP. It is neither a unary, nor a binary operator and cannot be disabled during the training. All other functions in the remaining layers are \cp-even. Since all terms in the output must contain the TP exactly once, it is \cp-odd by construction. We refer to this part $D_\mathrm{odd}(x)$. 
    \item Finally, we combine the two instances in a multiply layer and obtain $D(x) = D_\mathrm{even}(x) \cdot D_\mathrm{odd}(x)$ from which the optimal observable is constructed according to \cref{eq:tth_optimal_observable}.
\end{itemize}

First, we show in Fig.~\ref{fig:tth-cpodd-vars} the distributions of the two TPs which yield the best \cp-sensitivities at parton level and detector level, respectively. The momenta in \cref{eq:tth_cpodd_epsilon} are built from the Higgs momentum $p_h$, as well as even ($p_i + \bar{p}_i$) and odd ($p_i - \bar{p}_i$) combinations of the final state particles and their antiparticles. At parton level, the undecayed top quarks are included. This leads to 
\begin{align}
    \epsilon_{\mathrm{parton}} = \epsilon_{\mu\nu\rho\sigma}(p_t + p_{\bar{t}})^\mu(p_t - p_{\bar{t}})^\nu(p_l + p_{\bar{l}})^\rho(p_l - p_{\bar{l}})^\sigma
\end{align}
being the best TP on parton level (from which the $\Delta \phi_{ll}^{\ttbar}$ can be derived). This is expected because the top quarks transfer their polarization information to their decay products and the leptons hold the maximal spin analyzing power~\cite{Bigi:1986jk,Czarnecki:1990pe}. At the detector level (see Sec.~\ref{sec:ttH}),
\begin{align}
     \epsilon_{\mathrm{reco}} = \epsilon_{\mu\nu\rho\sigma}(p_b + p_{\bar{b}})^\mu(p_b - p_{\bar{b}})^\nu(p_l + p_{\bar{l}})^\rho(p_l - p_{\bar{l}})^\sigma
\end{align}
yields the highest \cp sensitivity. Here, we assume that the bottom and anti-bottom can be distinguished. This is experimentally very difficult. Our goal is, however, to give an optimistic estimate for the sensitivity of \cpo observables in $t\bar tH$ production. 

%--------------------------------
\begin{figure}[t]
    \centering
    \includegraphics[width=.48\textwidth]{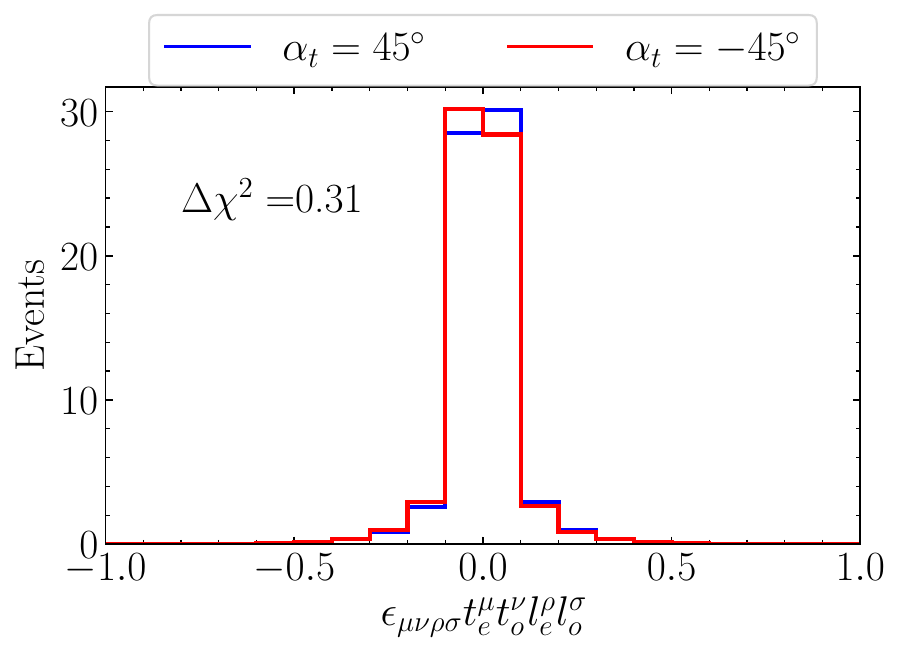}
    \includegraphics[width=.48\textwidth]{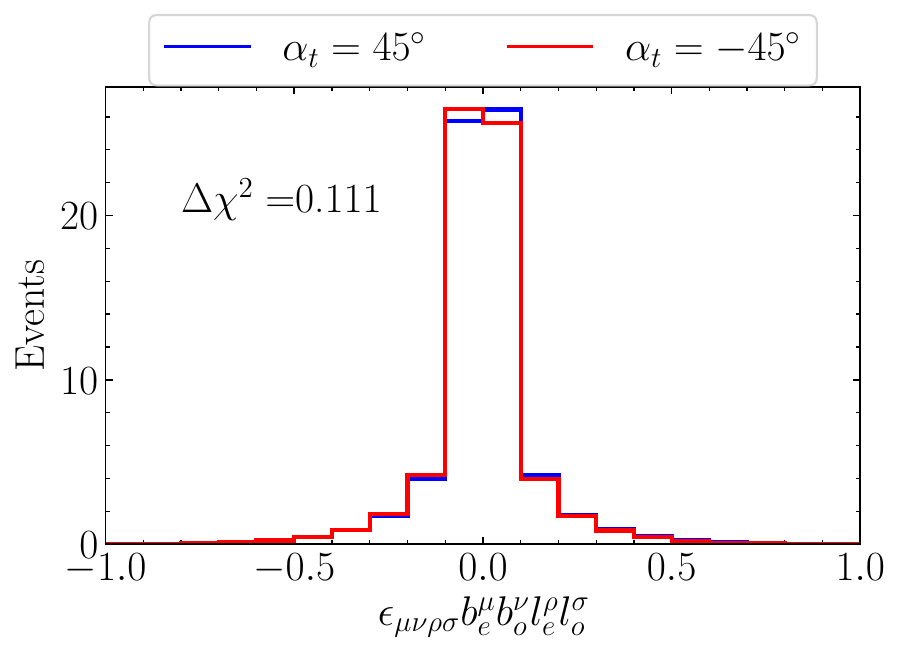}
    \caption{Distributions of $\epsilon_{\mu\nu\rho\sigma}(p_t + p_{\bar{t}})^\mu(p_t - p_{\bar{t}})^\nu(p_l + p_{\bar{l}})^\rho(p_l - p_{\bar{l}})^\sigma$ on parton level (left) and $\epsilon_{\mu\nu\rho\sigma}(p_b + p_{\bar{b}})^\mu(p_b - p_{\bar{b}})^\nu(p_l + p_{\bar{l}})^\rho(p_l - p_{\bar{l}})^\sigma$ on detector level (right) for $\alpha_t = \pm 45^\circ$.} 
    \label{fig:tth-cpodd-vars}
\end{figure}
%--------------------------------

%--------------------------------
\begin{figure}[b]
    \centering
    \includegraphics[width=.48\textwidth]{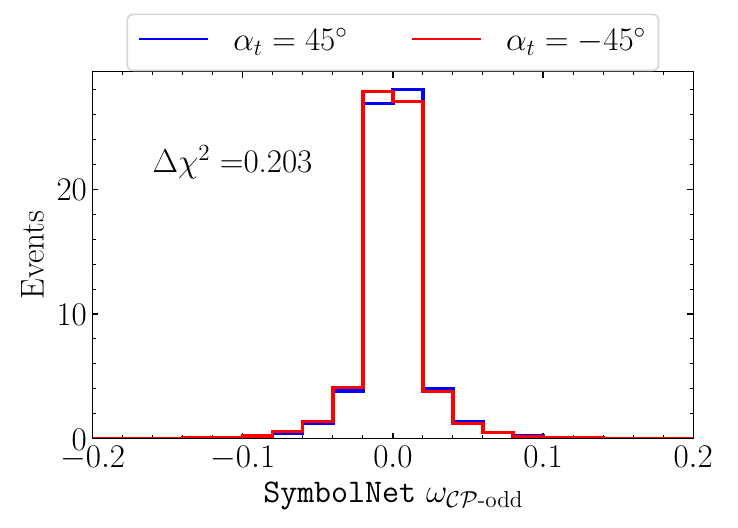}
    \caption{Distribution obtained from the equation predicted by \snet.} 
    \label{fig:tth-cpodd-snet}
\end{figure}
%--------------------------------

Next, we train \snet with the detector-level events to construct an optimal \cp-odd observable following the strategy outlined above. We use exactly one $V \to S$ and one $S \to S$ layer in both the \cp-even and \cp-odd part of the network, a learning rate of $0.005$ and a batch size of $256$. The result is shown in Fig.~\ref{fig:tth-cpodd-snet}. The final equation after sparsity training consists of
\begin{align}
    D_\text{odd}(x) =& \epsilon_{\mu\nu\rho\sigma} \left(0.32 p_{H} + 0.622 p_{b} - 0.164 p_{\bar{l}} + 0.634 p_{l}\right)^\mu \notag \\
    &\left(0.298 p_{H} + 0.196 p_{b} + 0.54 p_{\bar{b}} + 0.551 p_{\bar{l}} - 0.216 p_{l}\right)^\nu \notag \\
    & \left(+ 0.259 p_{H} + 0.448 p_{b} - 0.319 p_{\bar{l}} + 0.761 p_{l}\right)^\rho \notag \\
    &\left(0.176 p_{H} - 0.051 p_{b} + 0.488 p_{\bar{b}} + 0.772 p_{\bar{l}} - 0.462 p_{l} \right)^\sigma
\end{align}
and
\begin{align}
    D_\text{even}(x) =& \operatorname{Abs} \bigg( (- 0.123 p_{y,H} + 0.139 p_{y,b} + 0.159 p_{y,\bar{b}} - 0.375 p_{y,l} + 0.394 p_{z,H} \notag \\ 
    &+ 1.405 p_{z,b} - 0.134 p_{z,\bar{b}} - 0.627 p_{z,\bar{l}} + 0.803 p_{z,l} \notag \\
    & + 0.232 \Big\Vert - 2.715 p_{b} - 4.855 p_{\bar{b}} - 2.711 p_{\bar{l}} \Big\Vert_3 \notag \\
    &+ 0.016 \Big\langle \big(4.888 p_{H} - 3.127 p_{b} + 4.287 p_{\bar{b}} - 1.211 p_{\bar{l}} - 2.628 p_{l}\big) \notag \\
    & \times \big(- 2.516 p_{H} + 0.7615 p_{b} - 3.161 p_{\bar{b}} + 6.152 p_{\bar{l}} + 0.9251 p_{l} \big) \Big\rangle_3 \bigg)^{1/2}
    \; .
\end{align}
As the $\Delta \chi^2$ values of the distributions show, there is a big jump in sensitivity from a single TP to the output of \snet. Compared to the best parton-level TP, the best reco-level TP only reaches $36\%$ of the total sensitivity, while \snet reaches $65\%$. Over different training runs and hyperparameter settings, we observed that $D_\text{odd}(x)$ often collapses to a single, linear TP. We find equations containing non-linear terms in the TP to not lead to higher $\Delta \chi^2$ values.

%%%%%%%%%%%%%%%%%%%%%%%%%%%%%%%%%%%%%%%%%%%%%%%%%%%%%%%%%%%%%%%%%%%
%%%%%%%%%%%%%%%%%%%%%%%%%%%%%%%%%%%%%%%%%%%%%%%%%%%%%%%%%%%%%%%%%%%
\section{Test statistics for asymmetry}
\label{app:asymmetry}

The uncertainty per asymmetry bin can be obtained by propagating the Poisson errors for $N_{i}^+$ and $N_{i}^-$:
\begin{align}
    \sigma_{\mathcal{A}_i }
    &= 2 \sqrt{\frac{ \left(N_{i}^- \sigma_{N_{i}^+}\right)^2 +\left(N_{i}^+ \sigma_{N_{i}^-}\right)^2  }{(N_{i}^+ + N_{i}^- )^4} } = 2\sqrt{\frac{N_{i}^+ \cdot N_{i}^-}{(N_{i}^+ + N_{i}^-)^3}  }\;.
\end{align}
where the is obtained for $\sigma_{N_{i}^\pm} = \sqrt{N_{i}^\pm}$.

We then approximate the likelihood of $\mathcal{A}_i$ as a normal distribution $\mathcal{A}$ with mean $\mathcal{A}_i$ and width $\sigma_{\mathcal{A}_i}$. Thus, we can write the negative log-likelihood ratio between a BSM point and the SM as
\begin{align}
    q(\theta) = -2\log\frac{p(\{x\}|\theta)}{p(\{x\}|0)} = \sum_i\left[ \frac{\left(\mathcal{A}_i(\theta) - \mathcal{A}_i(0)\right)^2}{\sigma^2_{\mathcal{A}_i}(\theta)} + \log\frac{\sigma_{\mathcal{A}_i}^2(\theta)}{\sigma^2_{\mathcal{A}_i}(0)}\right]\;,
\end{align}
where we assume that the observed dataset $\{x\}$ follows the SM expectations --- \ie drawn from the likelihood $p(x|\theta)$.

Following Wilks' theorem, we assume the test statistic $q$ to follow a $\chi^2$ distribution. In our analysis, we exclude all bins with less than three events.

%%%%%%%%%%%%%%%%%%%%%%%%%%%%%%%%%%%%%%%%%%%%%%%%%%%%%%%%%%%%%%%%%%%
%%%%%%%%%%%%%%%%%%%%%%%%%%%%%%%%%%%%%%%%%%%%%%%%%%%%%%%%%%%%%%%%%%%

\section{Loss function for Collin-Soper angle reconstruction}
\label{app:loss}

One of the main challenges in most LHC analyses is to reconstruct the parton-level information from events obtained at the reconstruction level such that
\begin{align}
\label{eq:reco_likelihood}
    p_\text{reco}(x|\theta) = \int dx_\text{parton}~p(x_\text{reco}|x_\text{parton})~p(x_\text{parton}|\theta) \; .
\end{align}
This reco-level likelihood for a vector of parameters of interest $\theta$ depends on the parton level events via $p(x_\text{reco}|x_\text{parton})$. This conditional probability includes parton showering, hadronization, detector resolution, and any other effects that might appear. 

Our SR algorithms need to be able to approximate $p(x_\text{reco}|x_\text{parton})$. This raises the question of the optimal objective function for such a task. In a regression task with Gaussian errors on the fit parameters, we can show
\begin{align}
    -\log \loss_\text{Normal} \sim \sum_i \left( y_i - \hat{y}_i \right)^2 
\end{align}
for predicted (true) values $\hat{y}_i$ ($y_i$). Therefore, the MSE loss is optimal for solving such a problem. However, \cref{eq:reco_likelihood} includes much more than just a Gaussian smearing. For example, neutrino information is lost and jets may be wrongly reconstructed or identified. This means in practice that the training data contains outliers which can spoil the convergence of the SR algorithm. To avoid this, we can adapt the MSE loss function.

We obtained the best results using an ''inverse Gaussian loss'' 
\begin{align}
    \loss_\text{InvGaussian} = 1 - \exp{\left(-\frac{(y - \hat{y})^2}{2 \left( \frac{\max(y)}{\sigma} - \frac{\min(y)}{\sigma} \right)^2 }\right)}\;.
\end{align}
For $y\sim \hat{y}$, $\loss_\text{InvGaussian}$ collapses to the MSE loss, while for $|y - \hat{y}| \gg 0$ it approaches one, making it robust against outliers. $\sigma$ can be used to tune the loss to the expected range of predictions. 

%%%%%%%%%%%%%%%%%%%%%%%%%%%%%%%%%%%%%%%%%%%%%%%%%%%%%%%%%%%%%%%%%%%
\section{Formulas for the Collins-Soper angle}
\label{app:formulas}

Here, we present the formulas that were found by \pysr and \snet for the reconstruction of the CS angle $\cos \theta^*$. Individual components of the momenta are labeled by $E$, $p_x$, $p_y$, and $p_z$. Norms and dot products of 4-vectors correspond to the Minkowski norm and product, respectively. 

\subsubsection*{Scenario 1:}

\pysr:
\begin{align*}
    \cos \theta^*_{\pysr} =
    & \frac{p_{z,b} + p_{z,\bar{l}} + p_{z,\nu}}{\sqrt{\left(p_{x,b} + p_{x,\bar{l}} + p_{x,\nu}\right)^{2} + \left(p_{y,b} + p_{y,\bar{l}} + p_{y,\nu}\right)^{2} + \left(p_{z,b} + p_{z,\bar{l}} + p_{z,\nu}\right)^{2}}}
\end{align*}
\snet:
\begin{align*}
\cos \theta^*_{\snet} =
& \frac{1.006 p_{z,b} + 1.001 p_{z,\bar{l}} + 1.002 p_{z,\nu} - 1.027 p_{z,\bar{b}} - 1.027 p_{z,q} - 1.031 p_{z,\bar{q}}}{\Big\Vert 1.034 p_{b} + 1.022 p_{\bar{l}} + 1.024 p_{\nu} - p_{\bar{b}} - 1.007 p_{q} - 1.009 p_{\bar{q}}\Big\Vert_3}
\end{align*}

\subsubsection*{Scenario 2:}

\pysr:
\begin{align*}
\cos \theta^*_{\pysr} =
& \sin \bigg[1.41 \Big(p_{z,b} + 1.41 p_{z,\bar{l}} + 1.41 p_{z,\nu} - 1.41 p_{z,\bar{b}} - 1.41 p_{z,q} - 1.41 p_{z,\bar{q}}\Big) \Big/ \\
& \Big(E_{b} + E_{\bar{l}} + E_{\nu} + E_{\bar{b}} + E_{q} + E_{\bar{q}} \\
& + \sqrt{\left(- p_{y,b} - p_{y,\nu} + p_{y,\bar{b}} + p_{y,\bar{q}}\right)^{2} + \left(p_{x,b} + p_{x,\bar{l}} + p_{x,\nu} - p_{x,\bar{b}} - p_{x,\bar{q}}\right)^{2}} - 1.70\Big) \bigg]
\end{align*}
\snet:
\begin{align*}
\cos \theta^*_{\snet} =
& 1.006 \bigg( \operatorname{boost} \bigg[- 2.793 p_{b} - 2.811 p_{\bar{l}} - 2.793 p_{\nu} + 2.806 p_{\bar{b}} + 2.835 p_{q} + 2.815 p_{\bar{q}} \Big| \\
& - 2.443 p_{b} - 2.434 p_{\bar{l}} - 2.456 p_{\nu} - 2.45 p_{\bar{b}} - 2.429 p_{q} - 2.398 p_{\bar{q}} \bigg] \bigg)~\bigg|_z~\bigg/ \\ 
& \bigg\Vert \operatorname{boost} \bigg[- 2.793 p_{b} - 2.811 p_{\bar{l}} - 2.793 p_{\nu} + 2.806 p_{\bar{b}} + 2.835 p_{q} + 2.815 p_{\bar{q}} \Big| \\
& - 2.443 p_{b} - 2.434 p_{\bar{l}} - 2.456 p_{\nu} - 2.45 p_{\bar{b}} - 2.429 p_{q} - 2.398 p_{\bar{q}} \bigg] \bigg\Vert_3
\end{align*}

\subsubsection*{Scenario 3:}

\pysr:
\begin{align*}
\cos \theta^*_{\pysr} =
& \sin  \bigg[ \bigg(- 0.300 \sqrt{E_T^{\text{miss}}} \left(p_{z,b} + 2.22 p_{z,\bar{l}}\right) \Big(p_{y,\bar{l}} + p_{x,b} + p_{x,\bar{l}} - 1.99\Big) \\
& + p_{z,b} + p_{z,\bar{l}} - p_{z,\bar{b}} - p_{z,q} - p_{z,\bar{q}}\bigg) \bigg/ \\
& \bigg(E_T^{\text{miss}} + 0.949 E_{b} + 0.949 E_{\bar{b}} + E_{\bar{l}} + E_{q} + 0.949 E_{\bar{q}} - 1.12\bigg) \bigg]
\end{align*}
\snet: 
\begin{align*}
\cos \theta^*_{\snet} =
& - 1.079 \bigg( \operatorname{boost} \bigg[ - 4.831 p_{b} - 4.919 p_{\bar{l}} - 0.2614 p^{\text{miss}} + 4.752 p_{\bar{b}} + 4.827 p_{q} + 4.711 p_{\bar{q}} \Big| \\
& 4.506 p_{b} + 20.27 p_{\bar{l}} + 8.622 p_{\bar{b}} + 10.6 p_{q} + 8.935 p_{\bar{q}} \bigg] \\
& - 0.698 \tanh{\left(0.711 p_{b} + 0.821 p_{\bar{l}} + 2.083 p^{\text{miss}} - 0.717 p_{\bar{b}} - 0.575 p_{q} - 0.672 p_{\bar{q}} \right)}\bigg)~\bigg|_z~\bigg/ \\
& \bigg( \bigg\Vert \operatorname{boost} \bigg[- 4.831 p_{b} - 4.919 p_{\bar{l}} - 0.2614 p^{\text{miss}} + 4.752 p_{\bar{b}} + 4.827 p_{q} + 4.711 p_{\bar{q}} \Big| \\
& 4.506 p_{b} + 20.27 p_{\bar{l}} + 8.622 p_{\bar{b}} + 10.6 p_{q} + 8.935 p_{\bar{q}} \bigg] \bigg\Vert_3 \\
& + 1.039 \bigg\Vert \tanh{\left(0.711 p_{b} + 0.821 p_{\bar{l}} + 2.083 p^{\text{miss}} - 0.717 p_{\bar{b}} - 0.575 p_{q} - 0.672 p_{\bar{q}} \right)} \bigg\Vert_3^2 + 0.089\bigg)
\end{align*}

\subsubsection*{Scenario 4:}

\pysr:
\begin{align*}
\cos \theta^*_{\pysr} =
& \sin \bigg[ \bigg(- p_{z,\bar{b}} - p_{z,q} - p_{z,\bar{q}} + 1.06 \sqrt{0.888 E_T^{\text{miss}} + 1} \left(p_{z,b} + 1.24 p_{z,\bar{l}}\right)\bigg) \bigg/ \\
& \bigg(E_T^{\text{miss}} + 0.968 E_{b} + 0.826 E_{\bar{b}} + E_{\bar{l}} + p_{y,\bar{l}}^{2} + E_{q} + E_{\bar{q}} \\
& + 0.494 \left(- p_{x,b} - p_{x,\bar{l}} + p_{x,\bar{b}} + p_{x,\bar{q}}\right) - 1.48\bigg) \bigg]
\end{align*}
\snet:
\begin{align*}
\cos \theta^*_{\snet} =
& 1.074 \bigg( \operatorname{boost}\bigg[- 4.341 p_{b} - 4.71 p_{\bar{l}} - 0.2542 p^{\text{miss}} + 4.087 p_{\bar{b}} + 4.28 p_{q} + 4.133 p_{\bar{q}} \Big| \\
& - 4.103 p_{b} - 14.52 p_{\bar{l}} - 7.11 p_{\bar{b}} - 7.261 p_{q} - 6.894 p_{\bar{q}} - 0.327 \bigg] \\
& + 0.668 \tanh{\left(0.672 p_{b} + 0.723 p_{\bar{l}} + 1.949 p^{\text{miss}} - 0.624 p_{\bar{b}} - 0.625 p_{q} - 0.62 p_{\bar{q}} \right)}\bigg)~\bigg|_z~\bigg/ \\
& \bigg( \bigg\Vert \operatorname{boost} \bigg[- 4.341 p_{b} - 4.71 p_{\bar{l}} - 0.2542 p^{\text{miss}} + 4.087 p_{\bar{b}} + 4.28 p_{q} + 4.133 p_{\bar{q}} \Big| \\
& - 4.103 p_{b} - 14.52 p_{\bar{l}} - 7.11 p_{\bar{b}} - 7.261 p_{q} - 6.894 p_{\bar{q}} - 0.327 \bigg] \\
& + 0.162 \tanh{\left(0.672 p_{b} + 0.723 p_{\bar{l}} + 1.949 p^{\text{miss}} - 0.624 p_{\bar{b}} - 0.625 p_{q} - 0.62 p_{\bar{q}} \right)} \bigg\Vert_3 \\
& 0.906 \bigg\Vert \tanh{\left(0.672 p_{b} + 0.723 p_{\bar{l}} + 1.949 p^{\text{miss}} - 0.624 p_{\bar{b}} - 0.625 p_{q} - 0.62 p_{\bar{q}} \right)} \bigg\Vert_3^2 + 0.154\bigg)
\end{align*}

\subsubsection*{Scenario 5:}

\pysr:
\begin{align*}
\cos \theta^*_{\pysr} =
& \sin \bigg[ \bigg(1.156 p_{z,b} - \frac{1.156 p_{z,\bar{b}}}{E_T^{\text{miss}} \left(0.710 p_{x,\bar{b}} + p_{x,q} + p_{x,\bar{q}}\right) + 0.892} \\
& + 2.06 p_{z,\bar{l}} - 1.156 p_{z,q} - 1.312 p_{z,\bar{q}}\bigg) \bigg/ \\
& \bigg(E_{b} + E_{\bar{b}} - \sqrt{E_{\bar{l}}} + 2 E_{\bar{l}} + p_{x,\bar{l}}^{2} + E_{q} + E_{\bar{q}} + p_{y,\bar{q}}^{2} + \left(- p_{y,\bar{l}} + p_{y,q}\right) - 0.511\bigg) \bigg]
\end{align*}
\snet:
\begin{align*}
\cos \theta^*_{\snet} =
& - 0.928 \bigg( \operatorname{boost} \bigg[ 1.673 p_{b} + 2.866 p_{\bar{l}} - 1.666 p_{\bar{b}} - 2.495 p_{q} - 2.294 p_{\bar{q}} \Big| \\
& - 3.822 p_{b} - 5.112 p_{\bar{l}} - 3.246 p_{\bar{b}} - 3.755 p_{q} - 3.459 p_{\bar{q}} \bigg ] \\
& - 0.131 \tanh{\left(0.8275 p_{\bar{l}} - 0.3036 p_{\bar{q}} \right)} \bigg)~\bigg|_z~\bigg / \\
& \bigg\Vert \operatorname{boost} \bigg[1.673 p_{b} + 2.866 p_{\bar{l}} - 1.666 p_{\bar{b}} - 2.495 p_{q} - 2.294 p_{\bar{q}} \Big| \\
& - 3.822 p_{b} - 5.112 p_{\bar{l}} - 3.246 p_{\bar{b}} - 3.755 p_{q} - 3.459 p_{\bar{q}} \bigg]  \bigg\Vert_3
\end{align*}

\subsubsection*{Scenario 6:}

\pysr:
\begin{align*}
\cos \theta^*_{\pysr} =
& \sin \bigg[\bigg( 1.114 p_{z,b} + 2.143 p_{z,\bar{l}} - 0.858 p_{z,\bar{b}} - 0.426 p_{z,q} - 1.088 p_{z,\bar{q}} \bigg) \bigg/ \\
& \bigg( E_T^{\text{miss}} E_{q} + E_{b} + E_{\bar{b}} + 1.85 E_{\bar{l}} + E_{\bar{q}} \\
& - \left(p_{x,b} + p_{x,\bar{l}}\right) \left(p_{x,\bar{b}} + p_{x,q} + p_{x,\bar{q}}\right) - 0.863 - \frac{0.205 p_{z,q}}{\sqrt{E_{q}}} \bigg) \bigg]
\end{align*}
\snet:
\begin{align*}
\cos \theta^*_{\snet} =
& 0.971 \bigg( \operatorname{boost} \bigg[ - 1.521 p_{b} - 3.066 p_{\bar{l}} + 1.421 p_{\bar{b}} + 2.218 p_{q} + 2.043 p_{\bar{q}} \Big| \\ 
& - 2.845 p_{b} - 4.559 p_{\bar{l}} - 2.575 p_{\bar{b}} - 2.753 p_{q} - 2.676 p_{\bar{q}} \bigg] \\
& - 0.256 \tanh{\left(0.646 p_{\bar{b}} - 0.634 p_{\bar{l}} + 0.648 p_{q} \right)}\bigg)~\bigg|_z~\bigg/ \\
& \bigg( \bigg\Vert \operatorname{boost} \bigg[ - 1.521 p_{b} - 3.066 p_{\bar{l}} + 1.421 p_{\bar{b}} + 2.218 p_{q} + 2.043 p_{\bar{q}} \Big| \\
& - 2.845 p_{b} - 4.559 p_{\bar{l}} - 2.575 p_{\bar{b}} - 2.753 p_{q} - 2.676 p_{\bar{q}} \bigg] \bigg\Vert_3 \\
& + 0.774 \bigg\langle 0.071 \operatorname{boost} \bigg[- 1.521 p_{b} - 3.066 p_{\bar{l}} + 1.421 p_{\bar{b}} + 2.218 p_{q} + 2.043 p_{\bar{q}} \Big| \\
& - 2.845 p_{b} - 4.559 p_{\bar{l}} - 2.575 p_{\bar{b}} - 2.753 p_{q} - 2.676 p_{\bar{q}} \bigg] \\
& + \tanh{\left(0.646 p_{\bar{b}} - 0.634 p_{\bar{l}} + 0.648 p_{q} \right)} \times \\
& \tanh{\left(0.646 p_{\bar{b}} - 0.634 p_{\bar{l}} + 0.648 p_{q} \right)} \bigg\rangle_3 - 0.223 \bigg)
\end{align*}
%

%%%%%%%%%%%%%%%%%%%%%%%%%%%%%%%%%%%%%%%%%%%%%%%%%%%%%%%%%%%%%%%%%%%
%%%%%%%%%%%%%%%%%%%%%%%%%%%%%%%%%%%%%%%%%%%%%%%%%%%%%%%%%%%%%%%%%%%

\clearpage
\bibliography{tilman,bibliography}

\end{document}

%% file: incl_settings.tex
\usepackage[utf8]{inputenc} % input encodings (allow UTF-8 input)
\usepackage[T1]{fontenc} 	% selecting font encodings (use 8-bit T1 fonts)
\usepackage[english]{babel} % multilingual support (English language/hyphenation)

\usepackage[bitstream-charter]{mathdesign}
\usepackage{geometry} 		% flexible and complete interface to document dimensions
\usepackage{amsmath} 		% math package (American Mathematical Society)
\usepackage{mathtools} 		% math package (fixes various deficiencies of amsmath)
\usepackage{float} 			% floating objects such as figures and tables
\usepackage{graphicx} 		% enhanced support for graphics
\usepackage{tabularx} 		% tabulars with adjustable-width columns
\usepackage{booktabs} 		% professional-quality tables
\usepackage{color, xcolor} 	% foreground and background colour management
\usepackage{pdfpages} 		% inclusion of external multi-page PDF documents
\usepackage{extarrows} 		% extra arrows beyond those provided in amsmath
\usepackage{multirow} 		% create tabular cells spanning multiple rows
\usepackage{multicol} 		% intermix single and multiple columns
\usepackage{enumitem} 		% control layout of itemize, enumerate, description
\usepackage{xspace} 		% define commands that appear not to eat spaces
\usepackage{stackrel} 		% enhancement to the \stackrel command
\usepackage{tikz} 			% create PostScript and PDF graphics
\usepackage{braket} 		% Dirac bra-ket notation
\usepackage{bm} 			% access bold symbols in maths mode
\usepackage{tensor} 		% typeset tensors (tensor-style super- and subscripts)
\usepackage{slashed} 		% slash through characters (Feynman slash notation)
\usepackage{siunitx} 		% SI units package (typesetting values with units)
\usepackage{lastpage} 		% reference last page
\usepackage{cite} 			% improved citation handling
\usepackage[normalem]{ulem} % package for underlining
\usepackage{fontawesome} 	% access to web-related icons
\usepackage{tocloft} 		% control over the typography of the TOC, LOF, and LOT
\usepackage{titlesec} 		% interface to sectioning commands (various title styles)
\usepackage{doi} 			% create correct hyperlinks for DOI numbers
\usepackage{hyperref} 		% hypertext links (handel cross-referencing commands)
\usepackage[most]{tcolorbox} 					% coloured and framed text boxes
\usepackage[nameinlink, capitalize]{cleveref} 	% intelligent cross-referencing
\usepackage[nottoc, notlot, notlof]{tocbibind} 	% add references/index/contents to TOC
\usepackage[ruled, vlined]{algorithm2e} 		% floating algorithm environment
\usepackage{makecell}
\usepackage[makeroom]{cancel}
\usepackage{feynmf}
\usepackage{cancel}

\makeatletter
\def\BState{\State\hskip-\ALG@thistlm}
\makeatother

\makeatletter
\@ifundefined{pdfoutput}{}{\DeclareGraphicsRule{*}{mps}{*}{}}
\makeatother

\makeatletter
\DeclareRobustCommand*{\bfseries}{%
   \not@math@alphabet\bfseries\mathbf
   \fontseries\bfdefault\selectfont
   \boldmath
}
\makeatother

\hypersetup{
	pdftitle={Learned Uncertainties},
	pdfauthor={},
	colorlinks=true, 			% false: boxed links, true: colored links
	linkcolor={red!50!black}, 	% color of internal links (sections, pages, etc.)
	citecolor={blue!50!black}, 	% color of citation links (links to bibliography)
	urlcolor={blue!80!black} 	% color of URL links (external links)
} 
\DeclareSymbolFont{usualmathcal}{OMS}{cmsy}{m}{n}
\DeclareSymbolFontAlphabet{\mathcal}{usualmathcal}

\SetArgSty{textnormal}
\SetKwComment{Comment}{{\small\#}~}{}
\SetCommentSty{mycommfont}

\setitemize{itemsep=0pt, parsep=0pt} 				% adjust itemize environment
\setenumerate{itemsep=0pt, parsep=0pt} 				% adjust enumerate environment
\setitemize{itemsep=2pt,topsep=2pt,parsep=0pt,partopsep=0pt,leftmargin=*}
\setenumerate{itemsep=0pt,topsep=2pt,parsep=0pt,partopsep=0pt,labelindent=3pt,leftmargin=*}
\newlist{todolist}{itemize}{2}
\setlist[todolist]{label=$\square$}
\usepackage{pifont}

\usepackage{amsmath}
 
\usepackage{amsthm} 		% math package (typesetting theorems using AMS style)
\theoremstyle{definition}

\usetikzlibrary{arrows, shapes.geometric, arrows.meta, shapes, decorations.pathreplacing, fit, patterns, patterns.meta}

\definecolor{Rcolor}{HTML}{E99595}
\definecolor{Gcolor}{HTML}{C5E0B4}
\definecolor{Bcolor}{HTML}{9DC3E6}
\definecolor{Ycolor}{HTML}{FFE699}
\definecolor{Ycolor_light}{HTML}{FFF7DE}
\definecolor{Gcolor_light}{HTML}{F1F8ED}

\tikzstyle{expr} = [circle, minimum width=1.8cm, minimum height=1.8cm, text centered, align=center, inner sep=0, draw,font=\LARGE]
\tikzstyle{txt_huge} = [align=center, font=\Huge, scale=2]
\tikzstyle{txt} = [align=center, font=\LARGE, minimum height=1cm]
\tikzstyle{cinn} = [double arrow, double arrow head extend=0cm, double arrow tip angle=130, shape border rotate=90, inner sep=0, align=center, minimum width=2.1cm, minimum height=2.3cm, fill=Bcolor, draw,font=\LARGE]
\tikzstyle{cinn_black} = [cinn, minimum height=2.5cm, fill=black]
\tikzstyle{arrow} = [thick,-{Latex[scale=1.0]}, line width=0.2mm, color=black]
\tikzstyle{loss} = [rectangle, align=center,  minimum width=1.8cm, minimum height=1.5cm,fill=Rcolor,font=\LARGE, rounded corners]
\tikzstyle{xt} = [rectangle, align=center,  minimum width=5cm, minimum height=1.5cm,fill=Gcolor,font=\LARGE, rounded corners]
\tikzstyle{xts} = [rectangle, align=center,  minimum width=1cm, minimum height=1.5cm,fill=Gcolor,font=\Large, rounded corners]

\tikzstyle{embed} = [rectangle, rounded corners=0.3ex, minimum width=1.5cm, minimum height=1cm, text centered, align=center, inner sep=0, fill=Ycolor, font=\large, draw]

\tikzstyle{small_cinn} = [double arrow, double arrow head extend=0cm, double arrow tip angle=130, inner sep=0, align=center, minimum width=1.1cm, minimum height=0.5cm, fill=Bcolor, draw]

\tikzstyle{transformer} = [rectangle, rounded corners, minimum width=6cm, minimum height=2.4cm, font=\large, fill=Gcolor_light, draw]

\tikzstyle{attention} = [rectangle, rounded corners=0.3ex, minimum width=5.5cm, minimum height=1.2cm, align=center, fill=Gcolor, draw, font=\large]

%% file: incl_shortcuts.tex
\definecolor{red_cb}{HTML}{e41a1c}
\definecolor{blue_cb}{HTML}{377eb8}
\definecolor{green_cb}{HTML}{4daf4a}
\definecolor{purple_cb}{HTML}{984ea3}
\definecolor{orange_cb}{HTML}{ff7f00}

\definecolor{EmeraldGreen}{HTML}{1ea78d}
\definecolor{EnglishRed}{HTML}{b02427}
\hypersetup{colorlinks=true,urlcolor=EmeraldGreen,citecolor=EmeraldGreen,linkcolor=EnglishRed}

\newcommand{\snet}{\texttt{SymbolNet}\xspace}
\newcommand{\pysr}{\texttt{PySR}\xspace}

\newcommand{\ie}{\text{i.e.}\;}

\newcommand{\qqquad}{\qquad\quad}

\newcommand\one{\leavevmode\hbox{\small1\normalsize\kern-.33em1}}
\newcommand{\lag}{\mathscr{L}}

\newcommand{\loss}{\mathcal{L}} 	% loss value

\newcommand{\madgraph}{\textsc{MadGraph5\_aMC@NLO}\xspace}

\newcommand{\pythia}{\textsc{Pythia8}\xspace}
\newcommand{\fastjet}{\textsc{FastJet}\xspace}

\newcommand{\delphes}{\textsc{Delphes}\xspace}

\newcommand{\arXiv}[2][]{%
	\ifthenelse{\equal{#1}{}}%
	{\href{http://arxiv.org/abs/#2}{arXiv:#2}}%
	{\href{http://arxiv.org/abs/#2}{arXiv:#2~[#1]}}}

\newcommand{\gev}{\text{GeV}}

\newcommand{\ifb}{\ensuremath{\text{fb}^{-1}} }

\def\slashchar#1{\setbox0=\hbox{$#1$}           % set a box for #1
   \dimen0=\wd0                                 % and get its size
   \setbox1=\hbox{/} \dimen1=\wd1               % get size of /
   \ifdim\dimen0>\dimen1                        % #1 is bigger
      \rlap{\hbox to \dimen0{\hfil/\hfil}}      % so center / in box
      #1                                        % and print #1
   \else                                        % / is bigger
      \rlap{\hbox to \dimen1{\hfil$#1$\hfil}}   % so center #1
      /                                         % and print /
   \fi}

\newcommand{\tikznode}[2]{%
\ifmmode%
\tikz[remember picture,baseline=(#1.base),inner sep=0pt] \node (#1) {$#2$};%
\else
\tikz[remember picture,baseline=(#1.base),inner sep=0pt] \node (#1) {#2};%
\fi}

\def\mathswitchr#1{\relax\ifmmode{\text{#1}}\else$\text{#1}$\xspace\fi}
\def\mathswitch#1{\relax\ifmmode#1\else$#1$\xspace\fi}

\newcommand{\cp}{\ensuremath{{\mathcal{CP}}}\xspace}
\newcommand{\cpo}{\ensuremath{{\mathcal{CP}\text{-odd}}}\xspace}
\newcommand{\cpe}{\ensuremath{{\mathcal{CP}\text{-even}}}\xspace}
\newcommand{\ttH}{\ensuremath{t\bar t H}\xspace}
\newcommand{\ttbar}{\ensuremath{t\bar t}\xspace}
\newcommand{\ts}{\ensuremath{\theta^*}\xspace}

\newcommand{\cHWtil}{\ensuremath{c_{H\widetilde W}}\xspace}

\newcommand{\XGBoost}{\texttt{XGBoost}\xspace}